\documentclass[useAMS,usenatbib]{mn2e}

\usepackage{graphicx}
\usepackage{color}
\usepackage{amsmath}
\usepackage{lscape}
\usepackage{natbib}
\usepackage{subfiles}
\usepackage{amssymb}
\usepackage{pstricks}

\def\Ha{H${\alpha}$}

\newcommand{\apj}{ApJ}
\newcommand{\apjs}{ApJS}
\newcommand{\apjl}{ApJL}
\newcommand{\aap}{A{\&}A}

\newcommand{\mnras}{MNRAS}
\newcommand{\aj}{AJ}
\newcommand{\araa}{ARAA}
\newcommand{\pasp}{PASP}
\newcommand{\nat}{Nature}
\newcommand{\apss}{Ap\&SS}

\title[Red QSOs in the M$_{\rm BH}$-M$_{*}$ plane]{\center{The M$_{\rm BH}$ - M$_{*}$ relation for X-ray obscured, red QSOs at 1.2$<$z$<$2.6}}
\author[A. Bongiorno et al.]
{\parbox{\textwidth}{A. Bongiorno,$^{1}$\thanks{E-mail: \texttt{angela.bongiorno@oa-roma.inaf.it (OAR)}}
R. Maiolino,$^{2,3}$ 
M. Brusa,$^{4,5,6}$
A. Marconi,$^{7}$
E. Piconcelli,$^{1}$ 
A. Lamastra,$^{1}$
M. Cano-D\'iaz,$^{8,1}$ 
A. Schulze,$^{9,10}$
B. Magnelli,$^{11}$
C. Vignali,$^{4}$
F. Fiore,$^{1}$
N. Menci,$^{1}$
G. Cresci,$^{12}$
F. La Franca$^{13}$
A. Merloni,$^{6}$
}\vspace{0.4cm}\\
\parbox{\textwidth}{$^{1}$INAF-Osservatorio Astronomico di Roma, Via di Frascati 33, 00040, Monteporzio Catone, Rome, Italy.\\
$^{2}$ Cavendish Laboratory, University of Cambridge, 19 J. J. Thomson Ave., Cambridge CB3 0HE, UK.\\
$^{3}$ Kavli Institute for Cosmology, Madingley Road, Cambridge CB3 0HA, UK.\\
$^{4}$ Dipartimento di Astronomia, Universit\`a di Bologna, Via Ranzani 1, 40127, Bologna, Italy.\\
$^{5}$ INAF - Osservatorio Astronomico di Bologna, via Ranzani 1, 40127 Bologna, Italy.\\
$^{6}$ Max Planck Institut f\"ur Extraterrestrische Physik,  D-85478 Garching, Germany.\\
$^{7}$ Dipartimento di Astronomia e Scienza dello Spazio, Universit\`a degli Studi di Firenze, Largo E. Fermi 2, 50125 Firenze, Italy.\\
$^{8}$  Instituto de Astronom\'ia, Universidad Nacional Auton\'oma de M\'exico, A.P. 70-264,
04510, Mexico, D.F.\\
$^{9}$ Kavli Institute for the Physics and Mathematics of the Universe (Kavli IPMU, WPI), Todai Institutes for Advanced Study, the University of Tokyo, Kashiwa 277-8583.\\
$^{10}$ Kavli Institute for Astronomy and Astrophysics, Peking University, 100871 Beijing, China.\\
$^{11}$ Argelander-Institut f\"ur Astronomie, Universit\"at Bonn, Auf dem H\"ugel 71, 53121, Bonn, Germany.\\
$^{12}$ INAF - Osservatorio Astrofisico di Arcetri, Largo E. Fermi 5, 50125 Firenze, Italy.\\
$^{13}$ Dipartimento di Matematica e Fisica, Universita\`a degli Studi `Roma Tre', Via della Vasca Navale 84, I-00146 Roma, Italy
}}

\begin{document}

\date{Accepted Year Month Day. Received  Year Month Day; in original form Year Month Day}

\pagerange{\pageref{firstpage}--\pageref{lastpage}} \pubyear{2002}

\maketitle

\label{firstpage}

\begin{abstract}
We present near-infrared spectra, obtained with \texttt{SINFONI} and \texttt{XShooter} observations at ESO VLT, of nine dusty, red QSOs at $1.2<z<2.6$. The sources are hard X-ray detected, characterized by cold absorption (N$_{\rm H}>$10$^{21}$ - 10$^{22}$ cm$^{-2}$) and show a broad H$\alpha$ component in the NIR spectra. We complement this sample with twelve  additional sources taken from the literature with similar properties resulting in a total sample of 21 X-ray obscured, intermediate type (1.8-1.9), dusty reddened QSOs. From the broad H$\alpha$ line we have computed the BH masses through the virial formula and derived Eddington ratios. Moreover, from optical/IR multi-component SED fitting we have derived the stellar mass of their host galaxies and their SFRs. We find that most of the sources in our sample are hosted in starburst and main sequence star-forming galaxies with Eddington ratios $\lambda>0.1$. 

{We find a strong trend with the BH mass i.e. less massive objects are scattered below and above the local relation while the most massive ones are mainly located above it.} 
We also studied the evolution of these sources on the M$_{\rm BH}$-M$_{*}$ plane compared to a sample of optically blue type--1 QSOs and we find that obscured red QSOs show a ratio of M$_{\rm BH}$ to M$_{*}$ that increases with redshift which is consistent with or slightly lower  than what has been found for blue QSOs. 

These sources may represent the blow-out phase at the end of the rapid BH growth and immediately preceding the classical blue QSOs typically sampled in optical surveys. They in fact show evidence of outflows in the ionized gas component, but their BH has already fully formed.

\end{abstract}


\section{Introduction}
Tight scaling relations between the central BH mass and various properties of their host galaxies in the local Universe (velocity dispersion of the bulge component, $\sigma$, stellar masses, M$_*$, luminosity), have been discovered after the launch of the Hubble Space Telescope (HST) nearly 20 years ago. These correlations \citep{Magorrian1998,Ferrarese2000,Marconi2003, Haring2004}  have revolutionized the way we conceive the physical link between galaxy and AGN evolution, suggesting a possible physical coupling between BHs and host galaxies in their evolution.

The local scaling relations, although important in establishing the role of SMBHs in galaxy evolution, have been unable to uniquely determine the physical nature of the SMBH-galaxy coupling. A large number of theoretical models (some including feedback) have been proposed which can reproduce them reasonably well \citep{Silk1998,Granato2004,King2005}, but make different predictions for their redshift evolution. Constraining the high-z M$_{\rm BH}$-M$_{*}$ plane is thus of fundamental importance in discriminating between different models but it is unfortunately observationally complex. First of all, we have to rely on less-accurate, single-epoch virial mass estimators based on broad AGN emission lines. Moreover, in deriving the host galaxy properties,   we are limited by the presence of the bright nucleus.

At high redshift, most studies of the M$_{\rm BH}$-M$_*$ relationship have focused on optically blue type--1 broad line AGN (BLAGN) \citep[e.g.][]{Peng2006, Maiolino2007, Merloni2010, DeCarli2010,Jahnke2009,Salviander2013, Schramm2013} for which the UV broad lines (e.g. MgII and CIV, observed in the optical band at high-z) are detected and virial formulas can be used to derive BH masses from the AGN luminosity and the width of the broad lines \citep[see e.g.,][for recent reviews]{Peterson2013, Shen2013}. The emerging picture \citep[see][for a comprehensive picture]{Kormendy2013}  points towards a mild or absent evolution of the average black-hole-to-host-galaxy-mass ratio up to z$\sim$1 \citep{Jahnke2009,Cisternas2011b, Schramm2013} and a positive one for z$>$1 \citep[e.g.][]{Merloni2010,DeCarli2010, Bennert2011}.
In particular, putting together all available observations of AGN at z$\sim$0.1 - 7.1 that have reliable estimates of M$_{\rm BH}$, of the stellar masses of the host galaxies, and of bulge-to-disk ratios (for low-z objects only), \citet{Kormendy2013} finds that the M$_{\rm BH}$/M$_{*,bulge}$  evolves as (1+z)$^{\beta}$, with $\beta$=0.7 - 2.
The above results, have been interpreted either as evidence that black holes form faster, at earlier epochs, than their host galaxies, or as due to an evolution of the intrinsic scatter of the M$_{\rm BH}$-M$_*$ relation (or a combination of both effects). However, the sign of evolution seems to weaken at z$<2$ when considering the total galaxy mass for galaxies that are not bulge-dominated \citep{Jahnke2009, Kormendy2013}.

According to theoretical models \citep[e.g.][]{Lamastra2010,Menci2006,Hopkins2006a}, the positive evolution found for unobscured bright AGN at high redshift, can be interpreted as the fact that galaxy mergers, the driving mechanism for feeding BHs in bright QSOs, are more common at higher redshift, and are thus able to increase the BH accretion rate more rapidly than the star-formation rate at such redshifts. On the contrary, at lower redshift (z$<$1), for moderate-luminous, Seyfert-like objects with lower BH masses living in spiral-galaxy, secular galaxy evolution, occurring mostly in galaxy disks, would increase the stellar masses, with black hole accretion proceeding at a lower rate.
A radical alternative for explaining the M$_{\rm BH}$-M$_{\rm bulge}$ relation emerges from the statistics of mass averaging in galaxy mergers. This idea, first proposed by \citet{Peng2007}, has been elaborated using more realistic semi-analytic models of galaxy formation properly embedded in $\Lambda$CDM merger trees by \citet{Jahnke2011}. According to these simulations, the creation of the scaling relations can be fully explained by the
hierarchical assembly of BH and stellar mass through galaxy merging, from an initially uncorrelated distribution of BH and stellar masses in the early universe.

In contrast to the observational results mentioned above, \citet{Alexander2008} found that for AGN hosted in submm selected galaxies (SMGs) at 2$<z<$2.6, the growth of SMBHs actually lags that of the host stellar mass. 
The apparently contradictory behavior can be interpreted as a selection bias possibly affecting in the opposite way these studies. In fact, studying luminous AGN we are biased towards selecting the most massive SMBHs \citep[see][]{Lauer2007,Merloni2010,Schulze2011}, while a sample of SMGs will  be biased towards massive stellar hosts \citep[see][]{Lamastra2010} since the galaxies that are undergoing the most intense star formation at z$\sim$2 also appear to be massive galaxies at this epoch \citep[e.g.][]{Papovich2006}. 
In addition to possible bias in the sample selection, quasars with different properties may represent different phases in black hole and galaxy evolution.

Although SMGs are very interesting high-$z$ \citep[$\sim2$; e.g.,][]{Chapman2005}  objects with large star-formation activities (up to few 1000 M$_{\odot}\,$yr$^{-1}$; \citealt{Magnelli2012a}), they represent the most extreme class of sources of the entire star-forming galaxy population. New important constraints in understanding the physical nature of the AGN-galaxy coupling require the extension of the analysis of the BH-galaxy relation to the population of X-ray obscured, dust reddened QSOs. 
Current AGN-galaxy co-evolution models predict an early dust-enshrouded phase associated with rapid BH growth triggered by galaxy mergers \citep{Silk1998,DiMatteo2005,Hopkins2008}. Tidal interactions trigger both violent star formation and funneling of large amount of
gas into the nuclear region which feed and obscure the accreting SMBH \citep[e.g.,][]{Urrutia2008,Schawinski2010}.
During this phase, we expect the central source to be buried by the surrounding material and the AGN to likely appear as an optically type--2 AGN, X-ray obscured and red \citep[e.g.][]{Menci2008,Hopkins2008}.

So far, studies on this population are restricted to very few peculiar objects and the statistic is therefore not enough to draw a global picture. The observation of red QSOs is in fact more challenging compared to the blue ones. Measurements of BH masses for red QSOs at z$>$1 cannot be derived from optical spectroscopy: the UV broad emission lines, used to estimate the BH masses through virial estimators, are not detected due to dust extinction. For these sources, we can detect the AGN optical broad lines e.g., H$\alpha$, redshifted in the near-IR, since dust extinction is reduced with respect to the rest-frame UV lines \citep[see][]{Maiolino2006}. 

Until recently, there have been only a handful of published works on the evolution of the scaling relations between BHs and host masses for red QSOs: the above mentioned work of \citet{Alexander2008} based on  a sample of 4 QSOs detected in SMGs, the work from \citet{Sarria2010} based on three hard X-ray selected, obscured, red QSOs with broad H$\alpha$ emission line and the most recent work from \citet{Urrutia2012} based on a 2MASS+FIRST selected sample of 13 red QSOs at 0.4$<$z$<$1.
\citet{Sarria2010} found that the M$_{\rm BH}$/M$_*$ ratio at z=1-2 is in agreement with the local BH-galaxy relation, and lower than the average M$_{\rm BH}$/M$_*$ ratio observed in blue QSOs in the same redshift range. However, this result is still tentative since it is based on only three AGN.
On the contrary, \citet{Urrutia2012} found that most of the sources in their sample lie below the local relation and that objects below the local relation are the one accreting at high Eddington ratio ($\lambda=\frac{\rm L_{\rm bol}}{\rm L_{\rm edd}}>$0.3). 

Here we present new IR spectra obtained with \texttt{SINFONI} and \texttt{XShooter} at VLT of 9 X-ray obscured, red QSOs for which a broad H$\alpha$ component has been observed in the IR spectra. 
We complement these data with additional sources available in literature reaching  a final total sample of 21 X-ray obscured, intermediate type \citep{Osterbrock1981}, red QSOs with broad H$\alpha$ emission observed in the NIR. This represents the most extensive study so far for this class of targets. The aim of the paper is to study the M$_{\rm BH}$-M$_*$ scaling relation and its evolution for this class of sources by comparing them with the results obtained for optically blue QSOs.

The paper is organized as follows: in section
\ref{sec:sample} we will introduce our sample, starting from the new observations and proceeding with the samples taken from the literature and included in our analysis. 
In section \ref{sec:BH} and  \ref{sec:masses} we derive virial BH masses and host galaxy stellar masses. Star formation rates and Eddington ratios are presented in section \ref{sec:sfr_edd}.
Results on the M$_{\rm BH}$-M$_*$ plane are shown in section \ref{sec:scaling} followed by the comparison with the semi-analytical model prediction in section \ref{sec:model} . Finally, discussion and conclusions are presented in section \ref{sec:summary}. Unless otherwise stated, uncertainties are quoted at the 68\% (1$\sigma$) confidence level.

\section{The Sample}
\label{sec:sample}

This work is based on new near-IR observations (program 88.B-0316(A)  with \texttt{SINFONI} and 090.A-0830(A) with \texttt{XShooter} at the ESO VLT) of hard X-ray detected, obscured ($\rm N_{\rm H}\ge 10^{21}-10^{22} cm^{-2}$) AGN with red ($\rm R-K\gtrsim4.5$) colors which show a broad H$\alpha$ line, integrated with data collected from literature. The final sample consists of 21 objects, 9 of them are new targets while 
the remaining 12 are data taken from the literature as listed below.
The sample is summarized in Table \ref{tab:sample}.

\subsection{New observations}
\label{sec:sample_new}

Our observations were performed with \texttt{SINFONI} and \texttt{X-Shooter} at VLT.

\texttt{\underline{SINFONI Observations}:} We obtained near-IR spectra using \texttt{SINFONI} at VLT of eleven hard X-ray selected obscured ($\rm N_{\rm H}>10^{22} cm^{-2}$) red ($\rm R-K>4$) QSOs at $0.7<z<2.6$, selected from the \textit{Chandra Deep Field South} (CDFS) \citep[1Ms exposure,][]{Giacconi2002}. The sources were already spectroscopically classified as type--2 AGN in the optical band and had unambiguous spectroscopic redshifts \citep[GOODS-S,][]{Szokoly2004}. 
As our sample consisted of sources at different redshifts, observations were obtained in the $J$, $H$ and $K$ bands to detect the H$\alpha$ line. The resolutions in the three bands are respectively $R=2000$, $R=3000$ and $R=4000$.
The requested on-source integration time was 2.3 hours. 
However, of the 11 sources targeted, only 7 have been observed completely while 2 of them were observed for only 1.5 hours and the last 2 for 45 minutes. 
The typical seeing conditions during the observations were 0.5$'$$'$ - 1.1$'$$'$.

Data have been reduced by using the ESO-SINFONI pipeline. This pipeline performs the background subtraction, flat-fielding, the spectral calibration of each individual slice, and then reconstructs the cube. 
Out of these targets we detected a broad H$\alpha$ component for 4 of them (three of which are observed for the whole requested time while CDFS633 only for 45 minutes). Here we focus on these sources for which the H$\alpha$ broad component allowed us to infer their BH masses. 
If we restrict the sample to z$>$1 and $\rm f_{[2-10keV]}>1\times10^{-15} erg\, cm^{-2}s^{-1}$ we have seven objects, 4 of which with broad H$_{\alpha}$ detected. At the time of the observations, one source had a wrong redshift estimate and thus the H$\alpha$ line could not be detected while, for another one, the source was not at the center of the observed field. Therefore, we can conclude that out of 5 hard X-ray selected with $\rm f_{[2-10keV]}>1\times10^{-15} erg\, cm^{-2}s^{-1}$, obscured ($\rm N_{\rm H}>10^{22} cm^{-2}$) red ($\rm R-K>4$) QSOs at $1.0<z<2.6$, we detect a broad H$_{\alpha}$ line in 4/5 of the sample.

\texttt{\underline{XShooter Observations}:}  NIR spectra have been obtained using \texttt{XShooter} at VLT \citep{Vernet2011} of 10 X-ray obscured and luminous AGN at 1.25$<$z$<$1.72 selected  from the COSMOS field \citep[i.e. XMM-COSMOS sample,][]{Hasinger2007, Cappelluti2009, Brusa2010}. 
The parent sample consisted of $\sim150$ ``obscured QSOs'' candidates, selected
on the basis of their red optical colors ($R-K>4.5$) and high X-ray to optical flux ratio
and/or bright IR emission compared to the optical one \citep[see][Sec. 7 for details]{Brusa2010}. 
Among them, the observed sample consisted of the 6 K-brightest sources in the $1.25<z<1.72$ redshift range
with only photometric redshift available, plus 4 spectroscopically confirmed objects satisfying the
same redshift and magnitude selection (3 narrow-line objects and 1 broad line AGN). Compared to the CDFS sources, these sources
are brighter thus extending the range of luminosities covered by our sample to higher values.
The exposure times ranged from 1hr to 2hrs with typical seeing conditions of 0.5$'$$'$-1.0$'$$'$. Spectra have been taken with a resolution of R=5100 (0.9'' slit).
The reduction has been done using Reflex 2.4. A detailed
description of the full sample, data reduction and analysis can be found in Brusa et al. (2014). Here we focus
on 5 sources at $\rm L_{X}\sim10^{44}$ erg/s and $\rm N_{H}>10^{21-22} cm^{-2}$ for which a broad \Ha\ component has been
detected and hence BH masses could be computed. One of these sources is COSMOS2028, whose VLT/SINFONI spectra have been already published by \citet[][ULASJ1002+0137]{Banerji2012}. Given the lower resolution of their spectra (R=1500), \citet{Banerji2012} could not deblend the narrow and broad H$\alpha$ components.
The measured FWHM and consequently the BH mass is thus lower than the one obtained in this work.    
If we consider the z$>$1 and $\rm f_{[2-10keV]}>1\times10^{-14} erg\, cm^{-2}s^{-1}$ sources, we find that 5/6 of the observed sources have H$\alpha$ broad detected.

From these two observational campaigns, we thus find that, selecting hard X-ray obscured AGN with red colors, we are able to build a highly complete sample of objects ($\geq$80\%) with broad H$\alpha$ emission lines, above a certain flux.

Overall we have obtained a final sample of 9 sources with broad H$\alpha$ component. We fitted the line profile as explained in Sec. \ref{sec:fwhm} and derived the full width at half maximum of the line (FWHM). 
Stellar masses have been computed with a 2-component SED fitting procedure as explained in Sec. \ref{sec:masses}.

\subsection{Additional sources}
In order to increase the statistic of the study we complemented the new data with a collection of high-z sources taken from literature with similar properties.
Following is the list of included samples:\\

\begin{itemize}
\item \textit{3 QSOs from \citet{Sarria2010}} 

We include in our sample three X-ray obscured, red ($R-K>5$) QSOs from \citet{Sarria2010}. IR spectra of hard X-ray sources detected in the Hellas2XMM \citep{Fiore2003} and ELAIS-S1 survey \citep{Feruglio2008} have been taken with \texttt{SINFONI} and \texttt{Isaac}, and a broad H$\alpha$ line has been detected in three cases. 
This sample is the most similar to our new observations.   
For these objects, we re-analyzed the spectra after improving the reduction, sky subtraction and extraction.\\

\item \textit{6 QSOs hosted in SMGs from \citet{Alexander2008}}

We include in our work a sample of 6 SMGs that host X-ray identified obscured AGN with broad H$\alpha$ line from \citet{Alexander2008}. 
Four of these sources (the one with available stellar masses measured from \citealt{Hainline2011}) were already reported in \citet{Sarria2010}. Here we recomputed the stellar masses for all of the 6 sources using the SED fitting method as explained in Sec. \ref{sec:masses} after collecting the available photometry from literature. 
BH masses have been recomputed using the H${\alpha}$ FWHM measurements reported in \citet{Alexander2008} but differently from them, we used the de-absorbed X-ray luminosity \citep{Mushotzky2000,Manners2003,Alexander2003} instead of the continuum 5100\AA\ luminosity (see discussion in Sec. \ref{sec:BH}). The latter can in fact be absorbed and/or dominated by the stellar continuum. \\

\item \textit{2 QSOs hosted in ULIRGs from \citet{Melbourne2011}}

We include in our sample 2 X-ray obscured AGN hosted in ULIRGs observed with the Keck laser guide star adaptive optics (LGSAO) system and the OSIRIS Integral Field Spectrograph from \citet{Melbourne2011}. The targets (four in total) were selected from optical and MIR colors of the NOAO Deep Wide-Field Survey (NDWFS) of Bo\"otes to be Dust Obscured Galaxies \citep[DOGs,][]{Dey2008} and to have strong H$\alpha$ emission. 
BH masses have been recomputed using the formula derived in Sec. \ref{sec:BH} and since only two sources are detected in the 5ks {\it Chandra} X-ray observations of the Bo\"otes field \citep{Kenter2005}, we include only these two in our analysis.
As already pointed out by \citet{Melbourne2011} the use of X-ray luminosity leads to larger BH mass estimates, as expected by the fact that the rest-frame optical continuum is likely  absorbed by dust.
Stellar masses have been recomputed using the SED fitting technique as explained in Sec. \ref{sec:masses}, after collecting the available photometry from literature. \\

\item \textit{A QSO from \citet{DelMoro2009}}

We include in our sample also a QSO at $z=1.87$ from \citet{DelMoro2009}.
This object has been selected as the most extreme X-ray-to-optical flux ratio (EXOs)  sources amongst the sample of bright X-ray  selected EXOs obtained by cross-correlating 2XMMp and SDSS catalogs. New UKIRT optical/NIR photometry and MOIRCS IR spectroscopy have been presented in \citet{DelMoro2009}.
 X-ray luminosity and FWHM of the broad H$\alpha$ line, available in \citet{DelMoro2009}, have been used to compute the BH masses using the formula derived in Sec. \ref{sec:BH} while stellar masses have been derived from the K-band absolute magnitude. \citet{DelMoro2009} measures a K-band apparent magnitude of 19.9 in AB system,  
that converted in K-band absolute magnitude corresponds to M$_k$=-25.74. Assuming the relations between M$_*$ and L$_k$ for star-forming and quiescent galaxies found by \citet{Ilbert2010}, we estimated  logM$_*$=12.23$\pm$0.2\footnote{we note here that the average M$_*$/L$_k$ ratio of the sources of our sample for which the SED fitting procedure has been applied, fit well with this relation.}. \\

\end{itemize}

\begin{table}
\caption{Analyzed sample. $^{(a)}$ IDs from \citet{Giacconi2002}; $^{(b)}$ IDs from \citet{Cappelluti2009};  $^{(c)}$ ID from \citet{Feruglio2008};  $^{(d)}$ IDs from \citet{Mignoli2004}.}
\label{tab:sample}
\begin{tabular}{lccc}
\hline\hline
$ID$ & $RA$ & $DEC$ & $z$ \\
\hline
\multicolumn{4}{c}{New \texttt{SINFONI@VLT} observations of CDFS AGN}\\
\hline
  CDFSX57$^{\footnotesize{(a)}}$          &  03:32:12.9  & -27:52:36.9  &  2.561\\
  CDFSX153$^{\footnotesize{(a)}}$        &  03:32:18.3  & -27:50:55.3  &  1.536\\
  CDFSX531$^{\footnotesize{(a)}}$        &  03:32:14.4  & -27:51:10.9  &  1.544\\
  CDFSX633$^{\footnotesize{(a)}}$        &  03:31:50.4  & -27:52:12.2  &  1.374\\
\hline
\multicolumn{4}{c}{New \texttt{XShooter@VLT} observations of COSMOS AGN}\\
\hline
  COSMOS18$^{\footnotesize{(b)}}$       &  10:00:31.9  & +02:18:11.8  &  1.606\\
  COSMOS60053$^{\footnotesize{(b)}}$ &  10:01:09.2  & +02:22:54.7  &  1.582\\	   
  COSMOS175$^{\footnotesize{(b)}}$     &  09:58:53.0  & +02:20:56.4  &  1.530\\	   
  COSMOS2028$^{\footnotesize{(b)}}$   &  10:02:11.3  & +01:37:06.6  &  1.592\\	   
  COSMOS5321$^{\footnotesize{(b)}}$   &  10:03:08.8  & +02:09:03.5  &  1.470\\	   
\hline
\multicolumn{4}{c}{Re-analysis of the Sarria et al. (2010) data}\\
 \hline
  XMMES1\_460$^{\footnotesize{(c)}}$   &  00:36:41.5  & -43:20:38.1  &  1.748\\
  Ab2690\_29 $^{\footnotesize{(d)}}$     &  00:01:11.4  & -25:12:05.1  &  2.087\\
  BPM1627\_181$^{\footnotesize{(d)}}$  &  00:50:31.6  & -52:06:30.0  &  1.335\\
\hline
\multicolumn{4}{c}{Alexander et al. (2008)}\\
\hline
  A1236+6214                                & 12:36:35.5   & +62:14:24.0  &  2.015\\  
  A1237+6203                                & 12:37:16.0   & +62:03:23.0  &  2.053\\  
  A1312+4239                                & 13:12:15.2   & +42:39:00.0  &  2.555\\  
  A1312+4238                                & 13:12:22.3   & +42:38:14.0  &  2.560\\  
  A1636+4059                                & 16:36:55.8   & +40:59:14.0  &  2.592\\  
  A1637+4053                                & 16:37:06.5   & +40:53:13.0  &  2.375\\  
\hline
\multicolumn{4}{c}{Melbourne et al. (2011)}\\
\hline
  DOG2                                                    & 14:33:35.6   & +35:42:43.0  &  1.300\\	
  DOG3                                                    & 14:34:00.3   & +33:57:14.0  &  1.684\\	
 \hline
\multicolumn{4}{c}{Del Moro et al. (2009)}\\
\hline
  2XMMJ1232+2152                                 & 12:32:04.91  & +21:52:55.4  &  1.870\\ 
\hline\hline
\end{tabular}
\end{table}

\subsection{Comparison sample of optical type--1 blue QSOs from \citet{Merloni2010}}

As a comparison sample, we used a sample of $\sim$90 unobscured, optically type--1, blue QSOs \citep[$\rm R-K\sim$2-3,][]{Brusa2007} from the zCOSMOS survey in the redshift range 1$<z<$2.2,  presented in \citet{Merloni2010}. For consistency with our X-ray selection, only sources with XMM detections (82/89) have been included in the analysis. \citet{Merloni2010} computed their BH masses using the virial formula from \citet{McGill2008}. However, since we want to use these objects as comparison sample of the obscured one, we recomputed the BH masses  using the MgII line and the 3000\AA\ continuum luminosity as described in Sec \ref{sec:BH}, by adopting the same calibration method as for the obscured sample (eq. \ref{eq:2b}). For blue, unobscured QSOs, the continuum L$_{3000}$ is not expected to be severely affected by obscuration and can therefore be used in the virial equation.
For the stellar masses, since the method used by \citet{Merloni2010} is the same used in this paper, we do not recompute them but we use the ones provided in their Table 1.

\section{BH mass estimates}
\label{sec:BH}

BH masses have been obtained using the relation which combines the virial theorem with the radius-luminosity relation 
and provides  mass estimates which are dubbed as ``single epoch'', in contrast to the mass estimates from reverberation mapping observations \citep[e.g.][]{Peterson2004,Vestergaard2006}. The relation for single epoch mass estimates can be written as:
\begin{equation}
\label{eq:1}
\rm logM_{\rm BH}=\alpha + 2  log {\it{v_{1000}}}  + \beta log L_{44} 
\end{equation}
where L$_{44}$ is the luminosity estimator in units of 10$^{44}$ erg s$^{-1}$ and {\it v$_{1000}$} is the velocity estimator in units of 1000 km s$^{-1}$.
$\beta$ is the slope of the radius-luminosity relation which connects the size of the Broad Line Region (BLR) with the continuum luminosity \citep[e.g.][]{Kaspi2000, Bentz2009}. The normalization $\alpha$ is calibrated by assuming that the single epoch masses are in agreement with reverberation mapping masses \citep{Vestergaard2006} which, in turn, are in agreement with the $M_{\rm BH}-\sigma$ relation for normal galaxies \citep{Onken2004}. Such approach is motivated by 
 the unknown geometry and kinematics of the BLR (see, e.g., McLure \& Dunlop 2001, Onken et al. 2004). However, due to the different choices  adopted in the calibration, a large number of different relations are reported in the literature that relate black hole mass,  broad line width and continuum luminosity (e.g. \citealt{McGill2008,Kormendy2013}).
The key point for this paper is that the BH-galaxy scaling relation adopted to calibrate $\beta$ should be the same or be consistent with the scaling relation at z=0 used as a reference for studying the redshift evolution.

Here we adopt a relation which has been calibrated by directly imposing that single epoch masses agree with the local BH-galaxy scaling relations (Marconi et al. 2014, in preparation).
Briefly, single epoch virial products (i.e. BH masses for $\alpha=0$) have been estimated from H$\beta$ using the average spectra for the objects of the reverberation mapping database by \citet{Peterson2004} updated with the measurements from the most recent campaigns. In the case of MgII we have used the spectral measurements from International Ultraviolet Explorer (IUE) data performed by \citet{Wang2009}.
BH masses have been estimated from the stellar velocity dispersion \citep[see e.g.][ for a compilation of data]{Park2012}, or from the luminosity \citep{Bentz2009} of the host spheroid using the $M_{\rm BH}-\sigma$ and/or $M_{\rm BH}-L_V$ relations by \citet{Gultekin2009}.
The adopted relations can be expressed in the form

\begin{footnotesize}
\begin{equation}
\hspace{-1.4mm}\rm logM_{BH}=6.7+2 log\left(\frac{FWHM_{H\beta}}{10^{3} km \, s^{-1}}\right)+0.5log\left(\frac{\lambda L_{\lambda}(5100\AA)}{10^{44} erg \, s^{-1}}\right) 
\label{eq:2}
\end{equation}
\end{footnotesize}

\begin{footnotesize}
\begin{equation}
\hspace{-1.4mm}\rm logM_{BH}=6.6+2log\left(\frac{FWHM_{MgII}}{10^{3} km \, s^{-1}}\right)+0.5log\left(\frac{\lambda L_{\lambda}(3000\AA)}{10^{44} erg \, s^{-1}}\right) 
\label{eq:2b}
\end{equation}
\end{footnotesize}

\noindent
where FWHM refers to the H$\beta$ and MgII broad lines and L$_{\lambda}$(5100\AA) and L$_{\lambda}$(3000) represent the AGN continuum luminosity $\lambda L_\lambda$ at 5100\AA\ and 3000\AA, respectively. 
Uncertainties on the normalization, i.e. $\alpha$, are of the order of 0.1 dex (see Marconi et al. 2014, in preparation, for more details).
In the following, for the comparison with the $z=0$ relation, we will use as a reference the $\rm M_{ BH}-M_\star$ relation by \citet{Sani2011} because the sample adopted by those authors defines a  $\rm M_{BH}-\sigma$ relation consistent with that of \citet{Gultekin2009}.

To compute the BH masses of the objects in our sample, we used eq. \ref{eq:2} after converting H$\beta$ in H$\alpha$ using the correlation between H$\alpha$ and H$\beta$ widths found by \citet{Greene2005}. Moreover, since our sample is made of dust reddened QSOs, the continuum luminosity at 5100\AA\ is expected to be absorbed and/or strongly contaminated by stellar continuum. For this reason we replaced $\rm \lambda L_{\lambda}$(5100\AA) with the absorption corrected X-ray $\rm [2-10]keV$ luminosity using eq. (5) of \citet{Maiolino2007b}.
We obtain for the BH mass the following expression:
\begin{footnotesize}
\begin{equation}
\rm logM_{BH}=7.11+2.06\,log\,\frac{FWHM_{H\alpha}}{10^{3} km \, s^{-1}}+0.693\,log\,\frac{L_{[2-10keV]}}{10^{44} erg \, s^{-1}}
\label{eq:3}
\end{equation}
\end{footnotesize}

For our statistical analysis of the scaling relation evolution, we assign
to each black hole mass measurement an error given by the sum
of the statistical and systematic uncertainties.
The systematic uncertainty in the logM$_{\rm BH}$ determination has been estimated in 0.3 dex to account for the observed scatter in the virial relation itself, while in the computation of the statistical errors, we take into account the errors in the X-ray luminosity \footnote{an average 10\% error in the X-ray luminosities of all sources has been assumed in computing the BH masses.} and FWHM measurements, together with the uncertainties associated with the $\rm L_{[2-10keV]}-L_{5100}$ relation coefficient which contributes with 0.07 dex (in quadrature).

\begin{figure*}
\includegraphics[width=0.9\textwidth]{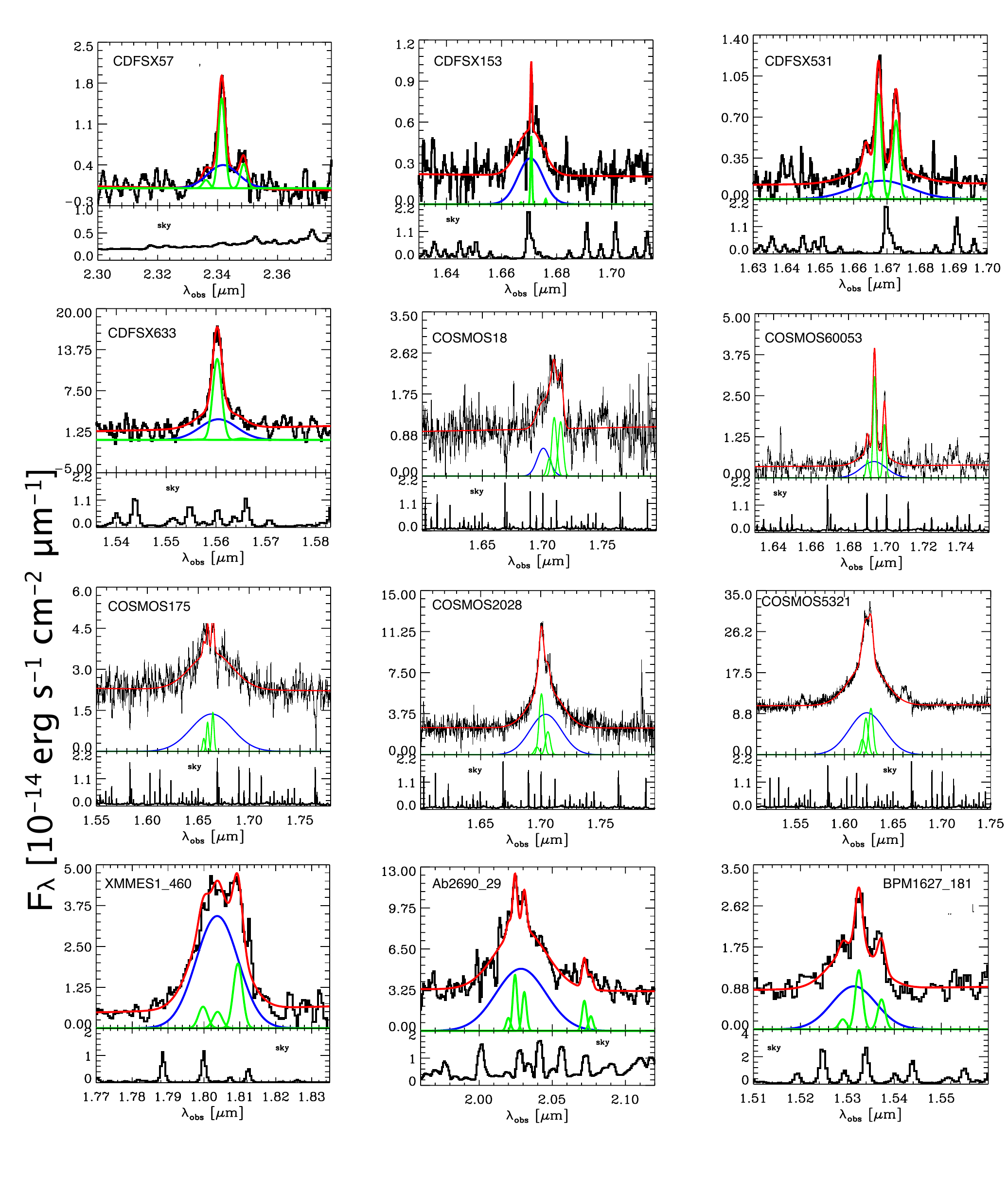}
\caption{Zoom-in showing the broad H$\alpha$ line of the near-IR spectra of the 9 X-ray obscured and red QSOs for which a broad H$\alpha$ component has been observed in XShooter and SINFONI. The last three spectra correspond to the objects already published in \citet{Sarria2010} for which we present the new reduction and fit. The blue line shows the broad H$\alpha$ Gaussian component while the green one corresponds to the narrow components for H$\alpha$ and [NII]. Finally the red line is the resulting fit. The bottom panel of each plot shows the sky spectrum.}
\label{fig:spectra_fit}
\end{figure*}

\subsection{FWHM measurements}
\label{sec:fwhm}

We measured the FWHM in the 9 new IR spectra presented here plus the three sources taken from \citet{Sarria2010} for which we re-analyzed the spectra after improving the sky subtraction and extraction. Following \citet{Sarria2010}, the broad H$\alpha$ component detected in the IR spectra has been fitted with a broad gaussian and one or more narrow gaussians accounting for the contribution of the narrow components of both H$\alpha$ and [NII]. The underlying continuum has been fitted with a linear component.
Regions affected by low S/N and/or bad sky residuals are not taken into account in the fit.
We linked the narrow components of H$\alpha$ and [NII] to have the same velocity shift and the same velocity dispersion. Moreover, the flux of the [NII]$\lambda$6548 line was linked to be one third of the [NII]$\lambda$6584 line.
The uncertainty in the width of the broad line is estimated by accounting for the correlation with the other components. 
A zoom of the 12 analyzed spectra around the region of the broad H$\alpha$ line, together with the line fit is shown in Fig. \ref{fig:spectra_fit}.
For the rest of the sources, we have taken the FWHM published values. 

Rest Frame FWHM values are reported in Table \ref{tab:parameters}.
As explained above, the measured FWHM of the H$\alpha$ broad component has been used, together with the de-absorbed X-ray $\rm [2-10]keV$ luminosity in Eq. \ref{eq:3} to compute the BH masses. 

\subsection{Intrinsic X-ray luminosity estimates}
\label{sec:lx}

Intrinsic X-ray [2-10keV] luminosities have been computed after correcting the flux for the absorption N$_{\rm H}$. 
When possible, N$_{\rm H}$ values have been derived directly by fitting the X-ray spectra. On the contrary, for sources detected with few counts, for which a proper X-ray spectral analysis cannot be performed, we have used the hardness ratio (computed as HR=(H-S)/(H+S) where S and H are the soft and hard band counts)  as a measure of obscuration. 
For the 9 sources for which new observations are presented, we have performed a fit of the X-ray spectra \citep[for details see][]{Mainieri2007,Mainieri2011, Merloni2014} for all but COSMOS60053, for which the quality of the X-ray data is low. For these sources, the N$_{\rm H}$ is typically constrained within 0.1 dex. This translates into uncertainties on the rest-frame X-ray fluxes (and therefore luminosities) from $<$5\%  for the brightest sources to $\sim$20\% for the faintest ones.
For the remaining sources, the N$_{\rm H}$ values have been taken from literature or derived from the hardness ratio (as specified in Table \ref{tab:parameters}) assuming a power-law X-ray spectrum with a typical photon index $\Gamma$ = 1.8 \citep{Piconcelli2005}.

Since X-ray luminosities are used, together with the FWHM, to compute the BH masses, it is important to be sure that the derived values are robust.  To this end, we have used the MIR/X-ray luminosity ratios \citep{Gandhi2009,Fiore2008,Lanzuisi2009} to check the consistency of the L$_{\rm X}$ with the whole SED of the source (see Sec. \ref{sec:masses}). 
We find that the agreement is good for all but 6 sources. In three of these  objects, the discrepancy is however relatively small (corresponding to a $\sim$0.3 dex difference in luminosity) and  can be ascribed to the large scatter present in the MIR/X-ray ratio used \citep{Gandhi2009}. On the contrary, for the remaining three objects (CDFSX633, A1236+6214 and A1237+6203), the discrepancy is larger i.e. $\sim$0.5 $-$ 0.6 dex in X-ray luminosity. In these cases, there is the possibility that the X-ray luminosity as derived by the X-ray data may be not reliable due to an underestimate of the N$_{\rm H}$ value caused by the low counts statistics. For these sources, we derived new intrinsic SED-derived X-ray luminosities and we used them in the computation of the BH masses (see next section). Intrinsic X-ray luminosities are reported in Table \ref{tab:parameters}.

\begin{figure*}
\includegraphics[width=0.8\textwidth]{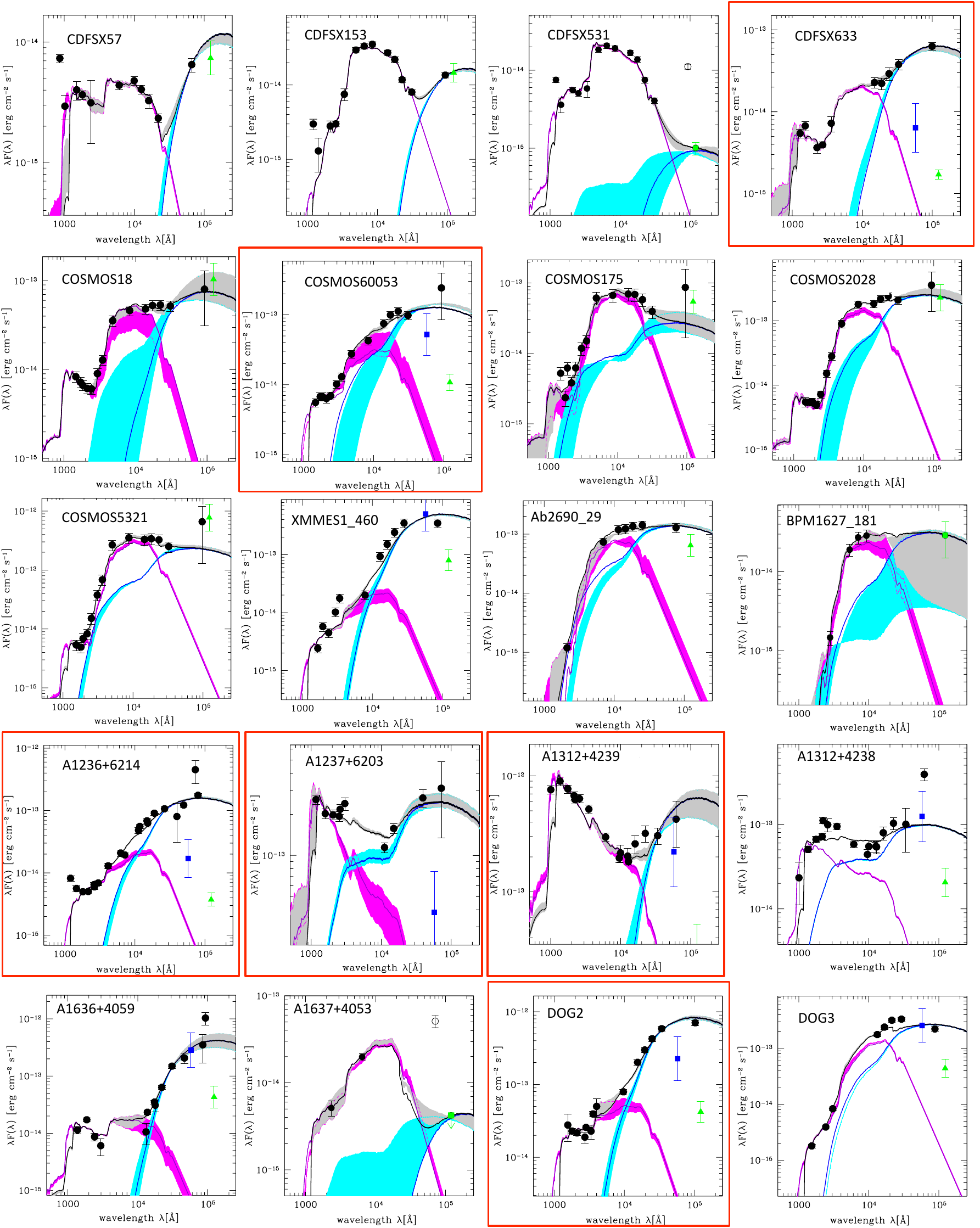}
\caption{SED fits for the 20 sources for which we derive stellar masses through multi-wavelength SED fitting. Black circles are rest-frame fluxes corresponding to the observed bands plotted rest-frame. Purple and blue lines correspond to the galaxy and the AGN template found as best-fit solution through the $\chi^2$ minimization, while the black line shows their sum.
Green triangles correspond to the rest-frame 12$\micron$ luminosity predicted from L$_{[\rm 2-10keV]}$ using the relation from \citet{Gandhi2009}. This point is taken into account (instead of the 24$\micron$ observed luminosity) in the fit of 3 sources in which the 24$\micron$ point is absent (BPM1627\_181) or clearly contaminated by star-formation emission(CDFSX531, A1637+4053). In the latter case, the observed 24$\micron$ point is showed as empty circle and the 12$\micron$ one as green square. Blue squares correspond to the 5.8$\micron$ luminosity predicted from L$_{[\rm 2-10keV]}$ using the relation from \citet{Lanzuisi2009}. We show this point only for the 10 objects in which the 12$\micron$ luminosity does not seem to agree with the whole SED.  Finally, red frames indicate the 6 cases in which both the rest-frame 12$\micron$ and the 5.8$\micron$ luminosities are in disagreement with the whole SED.} 
\label{fig:SED}
\end{figure*}

\begin{footnotesize}
\begin{table*}
\caption{Physical quantities for the selected sample. Stellar masses are computed using the Chabrier IMF. Objects marked with $^{(*)}$ are sources for which we report in the table the new corrected SED-derived L$_{\rm X}$ (see Sec. \ref{sec:masses}). The L$_{\rm X}$ as measured directly from X-ray data are kept in parenthesis, as well as the BH masses computed using these values. A 10\% error is associated to L$_{\rm X}$.  The reported N$_{\rm H}$ values are: 
(a) computed from X-ray spectral analysis; 
(b) computed from X-ray spectral analysis by \citet{Mainieri2007,Mainieri2011}; 
(c) computed from the HR by \citet{Merloni2014}; 
(d) taken from \citet{Sarria2010}; 
(e) computed from the HR in \citet{Alexander2003}; 
(f) computed from the HR in \citet{Mushotzky2000}; 
(g) computed from the HR in \citet{Manners2003}; 
(h) taken from \citet{Melbourne2011}; 
(i) computed from X-ray spectral analysis by \citet{DelMoro2009}. } 
\label{tab:parameters}
\begin{tabular}{lccccc}
\hline\hline
$\rm ID$ & $\rm L_{[2-10keV]}^{deabs}[erg/s]$ & N$_{\rm H}[\rm cm^{-2}]$ & $\rm FWHM_{H{\alpha}}[Km/s]$ & $\rm logM_{\rm BH}[M_{\odot}]$ & $\rm logM_{*}[M_{\odot}]$ \\
\hline
  CDFSX57                              &  $2.3\times 10^{44}$ & 2.0$\times 10^{23}$(a) & 1538$\pm$498 & 7.74$\pm$ 0.42 & $10.44^{+0.18}_{-0.11}$  \\ 
  CDFSX153                            &  $1.1\times 10^{44}$ & 3.0$\times 10^{23}$(a) & 1904$\pm$185 & 7.71$\pm$0.32 & $10.94^{+0.01}_{-0.01}$  \\
  CDFSX531                            &  $1.1\times 10^{43}$ & 2.0$\times 10^{23}$(a) & 3620$\pm$714& 7.59$\pm$0.45 & $10.59^{+0.01}_{-0.02}$  \\
  CDFSX633$^{(*)}$                 &  $7.2\times 10^{43}$ [$1.5\times 10^{43}$] & [3.0$\times 10^{23}$(a)] & 1665$\pm$557 & 7.47$\pm$0.43 [6.97]& $10.81^{+0.05}_{-0.09}$ \\ 
  COSMOS18                           &  $7.5\times 10^{44}$  & $3.2\times 10^{22}$(b) &   2162$\pm$531      & 8.41$\pm$0.38 &  $11.38^{+0.01}_{-0.21}$  \\
  COSMOS60053                     &  $9.5\times 10^{43}$  & $>1.0\times 10^{20}$(c) &   2776$\pm$391     & 8.00$\pm$0.33 &  $11.20^{+0.19}_{-0.41}$\\
  COSMOS175                         &  $3.8\times 10^{44}$  & $2.5\times 10^{21}$(b) & 8623$\pm$460 & 9.44$\pm$0.31 &  $11.55^{+0.12}_{-0.21}$  \\
  COSMOS2028                       &  $1.5\times 10^{45}$  & $7.9\times 10^{21}$(b) & 5423$\pm$146 & 9.44$\pm$0.31 &  $11.91^{+0.05}_{-0.03}$ \\
  COSMOS5321                       &  $3.8\times 10^{45}$  & $3.9\times 10^{21}$(b) & 7772$\pm$87      & 10.03$\pm$0.31 & $12.22^{+0.05}_{-0.01}$  \\
  XMMES1\_460                      &  $7.2\times 10^{44}$  & $3.2\times 10^{22}$(d)  & 2310$\pm$111& 8.45$\pm$0.31 & $10.53^{+0.31}_{-0.38}$    \\  
  Ab2690\_29                         &  $8.9\times 10^{44}$  & $6.3\times 10^{22}$(d)  & 5871$\pm$245& 9.35$\pm$0.31 & $12.00^{+0.03}_{-0.11}$      \\
  BPM1627\_181                     &  $1.6\times 10^{44}$  & $6.3\times 10^{22}$(d)  & 2173$\pm$330& 7.94$\pm$0.33 & $10.83^{+0.77}_{-0.21}$       \\
  A1236+6214$^{(*)}$            &  $3.5\times 10^{44}$  [$6.3\times 10^{43}$]& [$5.0\times 10^{23}$(e)] & 1600$\pm$200 & 7.90$\pm$0.33 [7.39]&    $10.46^{+0.01}_{-0.05}$ \\
  A1237+6203$^{(*)}$                   & 5.9$\times$ 10$^{44}$ [1.2$\times$10$^{44}$] & $3.2\times 10^{22}$(e) & 2400$\pm$500 & 8.43$\pm$0.36 [7.96]&    $10.54^{+0.25}_{-0.11}$ \\
  A1312+4239                  &  $7.9\times 10^{44}$   & $1.0\times 10^{21}{(f)}$ & 2500$\pm$500 & 8.55$\pm$0.35 &   $11.61^{+0.05}_{-0.23}$   \\
  A1312+4238                  &  $5.0\times 10^{44}$ & $1.0\times 10^{21}{(f)}$  & 2600$\pm$1000& 8.45$\pm$0.46 &    $10.59^{+0.06}_{-0.01}$   \\
  A1636+4059                  &  $1.0\times 10^{45}$ & $1.0\times 10^{22}{(g)}$  & 3000$\pm$400 & 8.79$\pm$0.33 &   $10.92^{+0.12}_{-0.45}$  \\
  A1637+4053                  &  $ <1.0\times 10^{44}$ &  ---  & 3300$\pm$1000& $<$8.18	 &   $11.35^{+0.22}_{-0.39}$  \\
  DOG2                              &  $2.1\times 10^{44}$ & $6.1\times 10^{21}$(h) &   2288$\pm$40    & 8.07$\pm$0.31  &  $10.47^{+0.41}_{-0.23}$	\\
  DOG3                              &  $3.9\times 10^{44}$ & $6.0\times 10^{20}$(h) &   6757$\pm$96    & 9.23$\pm$0.31  &  $11.91^{+0.02}_{-0.01}$  \\
  2XMMJ1232+2152                   &  $1.6\times 10^{46}$ & $2.0\times 10^{23}$(i) &   5280$\pm$331& 10.12$\pm$0.31 & $12.23^{+0.20}_{-0.20}$  \\

\hline\hline
\end{tabular}
\end{table*}
\end{footnotesize}

\section{Galaxy stellar mass estimates}
\label{sec:masses}

Stellar masses of the AGN host galaxies have been derived using a detailed two-component Spectral Energy Distribution (SED) fitting procedure \citep{Bongiorno2012} in which the observed optical-to-NIR SED is fitted with a large grid of models made from a combination of AGN and host-galaxy templates.

For the AGN component, we adopted the \citet{Richards2006SED} mean QSO SED derived from the study of 259 type--1 quasars with both Sloan Digital Sky Survey and Spitzer photometry. We allow for extinction of the nuclear AGN light by applying a SMC-like dust-reddening law \citep{Prevot1984} of the form: $A_{\lambda}/E(B-V) =1.39\lambda^{-1.2}_{\mu m}$. 
For the galaxy component we generated a library of synthetic spectra using the models of stellar population synthesis of \citet{Bruzual2003}.  
We assumed a universal initial mass function (IMF) from \citet{Chabrier2003} and we built 10 exponentially declining star formation histories (SFH) $SFR\propto e^{-t_{\rm age}/\tau}$ with
e-folding times, $\tau$, ranging from 0.1 to 30 Gyr and a model with
constant star formation.  
For each of the SFHs, the SED was generated for a grid of 13 ages ranging from 50 Myr to 9 Gyr, subject only to the constraint that 
the age should be smaller than the age of the Universe at the redshift
of the source. Moreover,  dust extinction of the galaxy component has been taken into account using the Calzetti's law \citep{Calzetti2000}, which is the most used attenuation curve in
high-redshift studies.

This technique, applied to data-sets with a wide multi-wavelength optical-to-NIR coverage, which is the case for most of our objects, allow us to decompose the entire spectral energy distribution into a nuclear AGN and a host galaxy components and to derive robust measurements of the host galaxy properties, i.e. stellar mass. For more details we refer the reader to \citet{Bongiorno2012}.

In the sample considered here, we used this technique for all but one of the 21 sources. The result of the SED fitting procedure applied to our data is shown in Fig. \ref{fig:SED} and the values of the host stellar masses, using a Chabrier IMF, are reported in Table \ref{tab:parameters}. In few cases, the most extreme blue data point show an excess compared to the whole SED and to the model fit. The excess corresponds to the rest-wavelength of the Ly$\alpha$ emission line and is not taken into account in the fitting procedure.
For the remaining source (2XMMJ1232+2152), no multi-wavelength photometry was available and hence our estimated stellar mass has been computed from the absolute K-band magnitude.

As in the 24$\micron$ flux measures there could be a strong contribution from star formation, we also computed the AGN 12$\micron$ fluxes derived from the X-ray luminosities using the \citet{Gandhi2009} relation to constrain the AGN-only emission. We find that in some cases these two quantities (12$\micron$ and 24$\micron$ fluxes) are indeed discrepant. 
In 2 cases (i.e., CDFSX531, A1637+4053) we have used the AGN 12$\micron$ point derived from the X-ray luminosity to constrain the AGN emission since the whole SED shape was, in terms of $\chi^2$, sligthly better constrained. Morever, for BPM1627\_181, no 24$\micron$ observations were available and thus the 12$\micron$ point was used.
In these three cases, the observed 24$\micron$ point is shown in Fig. \ref{fig:SED} as an open circle and indicates that the given stellar mass has been computed anchoring the AGN template to 12$\micron$ which is indeed relative only to the nuclear emission. 

It is worth noting that, for the source A1637+4053, the scarse observational data-points make the fit of this source, and thus M$_{*}$ poorly constrained. Moreover, the X-ray luminosity of this source is an upper limit and thus it is its M$_{\rm BH}$. 

In other cases (10/20, i.e., CDFSX633, COSMOS60053, XMMES1\_460, A1236+6214, A1237+6203, A1312+4239, A1312+4238, A1636+4059, DOG2, DOG3) the 12\micron\ luminosity derived from the de-absorbed X-ray luminosity is too low to fit the global SED shape. The discrepancy could be ascribed to effects of variability of the X-ray emission between the times of the X-ray and MIR observations. However, a detailed analysis of the light curve of the XMM-COSMOS and CDFS samples does not reveal any sign of  remarkable variability in their X-ray fluxes \citep{Lanzuisi2014,Young2012}. 
However, as already pointed out by \citet{Lanzuisi2009}, the relation between X-ray and MIR luminosity found locally by \citet{Gandhi2009} for optically/X-ray bright Seyfert--2 galaxies, does not always fit higher redshift and higher MIR luminosity, red obscured sources (i.e. Extreme DOGs, EDOGs defined to have F$_{24\micron}$/F$_{\rm R}>$2000 and F$_{24\micron}>$1.3mJy).
In particular, \citet{Lanzuisi2009} \citep[but see also][]{Fiore2008,Maiolino2007} find that X-ray selected QSO2 follow the local relation at all redshifts and luminosities while dust-absorbed powerful (ULIRG-like) nuclei are characterized by extreme MIR/X-ray luminosity ratios. For these sources a flatter relation is derived (which however shows a large scatter) that, for a given L$_{[2-10keV]}$, predicts an higher MIR luminosity  compared to the local one. 
For the sources for which the Gandhi relation does not predict a MIR luminosity in agreement with the whole SED we thus applied also the \citet{Lanzuisi2009} relation\footnote{The same result would be  obtained using the relation derived by \citet{Fiore2008} since, in the luminosity range spanned by the analyzed sample, the two relations are very similar.}. The new MIR luminosities predicted in this way are shown with blue squares in the SED fit in Fig. \ref{fig:SED} of the 10 sources listed above. 
For 4 of these cases (XMMES1\_460, A1312+4238, A1636+4059 and DOG3), the MIR luminosity predicted by the Lanzuisi relation are in agreement with the one expected by the whole SED. For the remaining 6 (CDFSX633, COSMOS60053, A1236+6214, A1237+6203, A1312+4239, DOG2) a discrepancy is still visible. 
As explained in Sec. \ref{sec:lx}, for the three sources for which the discrepancy corresponds to a difference of the order of 0.5 - 0.6 dex in the X-ray luminosity (CDFSX633, A1236+6214 and A1237+6203), we computed new intrinsic X-ray luminosities assuming the MIR luminosity as derived by the SED\footnote{Before doing that we double-checked, with a a different SED code which includes the starburst component, that the discrepancy found cannot be ascribed to SF contamination to the 24$\micron$ photometric point.} and applying the relation found by \citet{Lanzuisi2009}. 
These SED-derived L$_{\rm X}$ of these three sources have been used to  compute the BH masses and are reported in Table \ref{tab:parameters}. The original L$_{\rm X}$ values obtained from the X-ray data analysis (see Sec. \ref{sec:lx}) are also listed in Table \ref{tab:parameters} in parentheses.

\begin{table}
\caption{SFRs (Chabrier IMF) in M$_{\odot}$/yr and Eddington ratios $\lambda$ of the analyzed sample. $band$ indicates which data we used to compute the SFR. }
\label{tab:sfr_edd}
\begin{tabular}{cccc}
\hline
$\rm ID$ & $\rm log(SFR)$ & $\rm band$ & $\lambda$ \\
\hline
  CDFSX57                             & $<$2.56	             &  PEP &  1.38\\
  CDFSX153                           & $<$1.95              &  PEP &  0.56\\
  CDFSX531                           & 1.82$\pm$0.07   &  PEP &  0.038\\
  CDFSX633                           & $<$1.92              &  PEP & 0.56 \\
  COSMOS18                         &	 $<$2.21              & PEP  &  1.31\\
  COSMOS60053                   &	 2.87$\pm$0.01   &  PEP &  0.24\\
  COSMOS175                       &	 $<$2.14              &  PEP &  0.05\\
  COSMOS2028                     &	 2.44$\pm$0.01   &  PEP &  0.03\\
  COSMOS5321                     &	 2.36$\pm$0.01   &  PEP &  0.15\\
  XMMES1\_460                    &	 2.41$^{+0.14}_{-0.01}$ &  SED &  1.15\\ 
  Ab2690\_29                       &	 $<$0                   &  SED &  0.18\\
 BPM1627\_181                    & $<$0                   &  SED &  0.54\\
  A1236+6214                     &	 2.73$\pm$0.01   &  PEP &  1.66 \\
  A1237+6203                     &	 2.48$\pm$0.19   &  Scuba &  1.0\\
  A1312+4239                     &	 2.35$\pm$0.14   &  Scuba &  1.00\\
  A1312+4238                     &	 2.09$\pm$0.18   &  Scuba &  0.75\\
  A1636+4059                     &	 3.22$\pm$0.12   &  Scuba &  0.74\\
  A1637+4053                     &	 3.44$\pm$0.16   &  Scuba &  0.16\\
  DOG2                                 &	 2.56$\pm$0.10   &  SED    &  0.58\\
  DOG3                                &	 2.22$\pm$0.10   &  SED    &  0.09\\
  2XMMJ1232+2152            &   -99.0 	             &   ---   &  0.46\\
\hline
\end{tabular}
\end{table}

\section{Host Galaxy's star formation rates and BH Eddington ratios}
\label{sec:sfr_edd}

In this section, we study and compare the rate at which host galaxies accrete in relation with their BHs. To this end, we have derived the galaxy star formation rate and BH Eddington ratio for each  object in the sample. 
In general, the most reliable way to compute the galaxy's SFR is adding up the rate of unobscured star formation (emitted in the UV) with the
obscured one re-emitted in the (mid- and far-) IR by dust (SFR$_{\rm UV+IR}$). However, for  bright IR sources, SFR$_{\rm UV}$ is negligible compared to the absorbed one and we can simply assume $\rm SFR_{TOT} \sim SFR_{IR}$. For this reason, mid and far-IR measurements of cold dust
heated by star-forming processes are in principle the most reliable way to measure the galaxy's SFR. On the one hand, in case of bright QSOs, the AGN contamination in the IR band can be not negligible and thus the derived SFRs slightly overestimated \citep[see e.g.][]{Bongiorno2012,Mullaney2012}. 
On the other hand, the optical/IR SED fitting procedure relies only on the UV emission measurements, which traces the unobscured SF and takes into account the obscured component using the dust correction factor computed from the full SED. 
For the objects in our sample we estimated the host star formation rates using submm/IR data when available and the SED fitting procedure otherwise. 
The SFR of our sources are inferred from their infrared luminosities using the best set of mid/far-infrared/submm data available. For sources with PACS detections, IR luminosities are derived by fitting their far-infrared flux densities with the SED templates of \citet{Dale2002}. For sources with no far-infrared detection but with a submillimeter detection (i.e., from SCUBA), infrared luminosities are derived from their SCUBA flux densities using the L$_{\rm IR}$-submm relation of \citet{Magnelli2012a}. 
From these IR luminosities, we infer SFRs assuming SFR $\rm [M_{\odot} yr^{-1}]=10^{-10}\times L_{\rm IR}\,[L_{\odot}]$ \citep{Kennicutt1998} for a \citet{Chabrier2003} IMF.

\begin{figure}
\includegraphics[width=0.45\textwidth]{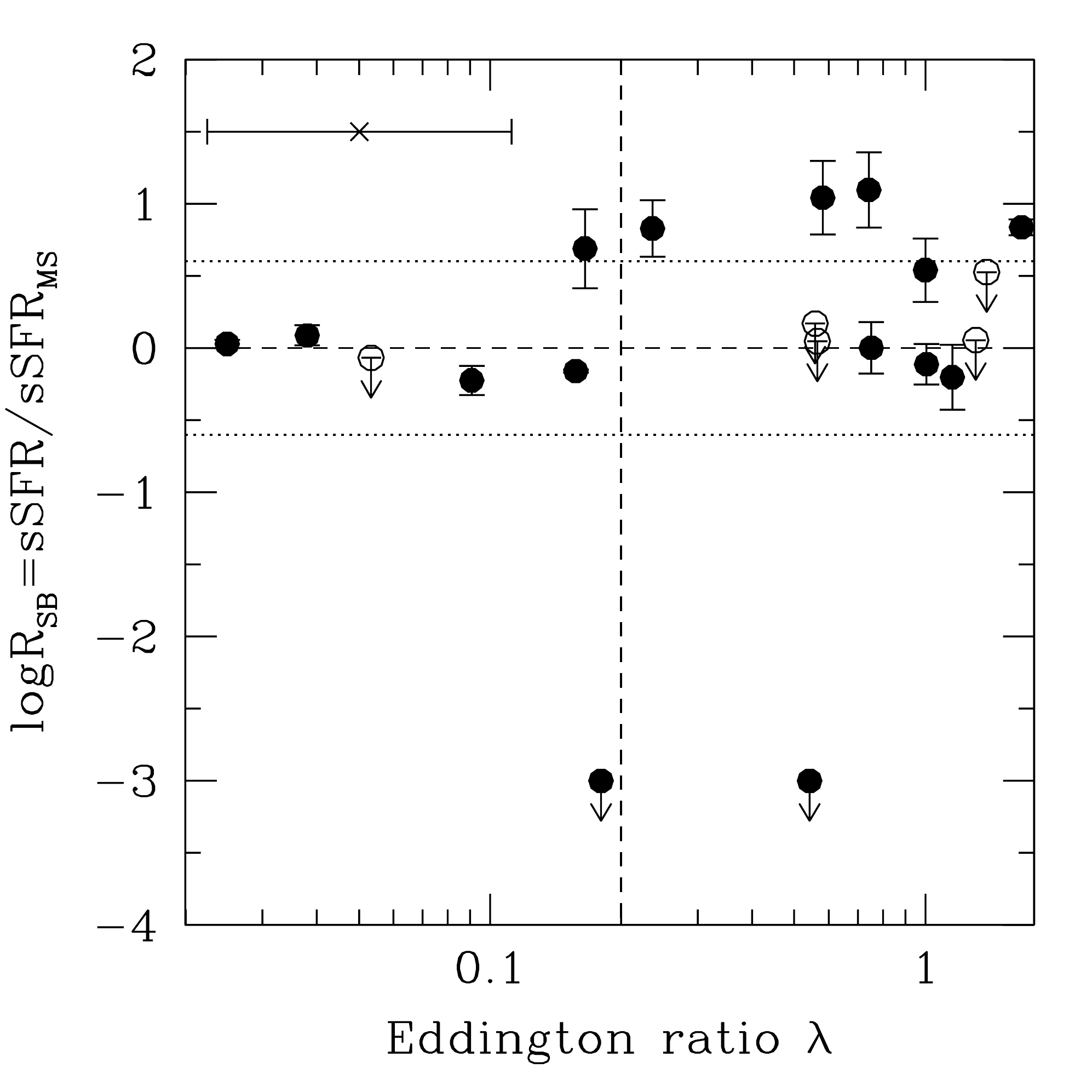}
\caption{Starburstiness R$_{\rm SB}$ = sSFR/sSFR$_{\rm MS}$ of the host galaxies versus the BH Eddington ratio. The sSFR values are normalized assuming the best-fit of the galaxy main sequence as a function of redshift (horizontal dashed line) obtained by \citet{Whitaker2012}. 
Dotted lines trace 4 times above and below the main sequence: the three delimited regions are used to define a galaxy to be starburst, main sequence or  quiescent \citep{Rodighiero2011}.}
 SFRs have been derived from IR/or submm fluxes when available and from SED-fitting otherwise (see Table \ref{tab:sfr_edd}). Open circles correspond to the cases for which we have only an upper limit on the SFR while filled circle with an arrow correspond to the objects with log(SFR) $<$0. The cross with the errorbar in the upper left corner shows the average error on the $x$-axis which corresponds to an error of $\sim$0.35 dex in $\lambda$.
\label{fig:scaling_z_sfr_edd}
\end{figure}

For quantifying the BH accretion, we derived the Eddington ratio, defined
as the ratio between the AGN bolometric luminosity and the Eddington luminosity.:
\begin{equation}
\rm \lambda_{\rm Edd} = \frac{L_{\rm bol}}{L_{\rm Edd}} = \frac{L_{\rm bol}\,[erg\,s^{-1}]}{1.3 \times 10^{38} \times M_{\rm BH}/M_{\odot}}
\end{equation}
\noindent
where L$_{\rm bol}$ have been derived from the X-ray luminosities using the bolometric correction derived from \citet{Marconi2004}.
Galaxy's SFRs and BHs Eddington ratios $\lambda$ are reported in Table \ref{tab:sfr_edd}. 

Figure \ref{fig:scaling_z_sfr_edd} shows the starburstiness R$_{\rm SB}$ = sSFR/sSFR$_{\rm MS}$ (sSFR$_{\rm MS}$ being the reference sSFR=SFR/M$_*$ of the galaxies in the main sequence) of the host galaxies versus the BH Eddington ratio. Open circles correspond to the cases for which we have only an upper limit on the SFR while filled circles with an arrow to the objects with log(SFR)$<$0.
The starburstiness R$_{\rm SB}$ \citep{Elbaz2011,Lamastra2013} measures the excess or deficiency in sSFR of a galaxy in terms of its distance from the galaxy main sequence (sSFR$_{\rm MS}$). For this comparison, we used the best-fit of the galaxy main sequence as a function of redshift obtained by \citet{Whitaker2012}.

Following \citet{Rodighiero2011}, we define a galaxy to be starburst if its SFR is above 4 times the main sequence (R$_{\rm SB}>$4),  quiescent if it is below 4 times the main sequence (R$_{\rm SB}<$1/4) and main sequence if 1/4$<R_{\rm SB}<$4 \citep[see also][]{Sargent2012, Lamastra2013a}.
Moreover, we divided the sample in high accreting BHs ($\lambda>$0.2) and low accreting BHs ($\lambda<$0.2; see Fig. \ref{fig:scaling_z_sfr_edd}).

We find that most (90\%) of our sources are hosted in starburst or main sequence star-forming galaxies with BHs accreting at $\lambda>$0.1 (80\%). 
No clear trend is visible between BH accretion and host galaxy's SFR. However, we note that starburst galaxies are preferentially found (all but one) to host high-Eddington ($\lambda>$0.2) AGN.

\section{The M$_{\rm BH}$ - M$_{*}$ scaling relation}
\label{sec:scaling}

\begin{figure}
\includegraphics[width=0.48\textwidth]{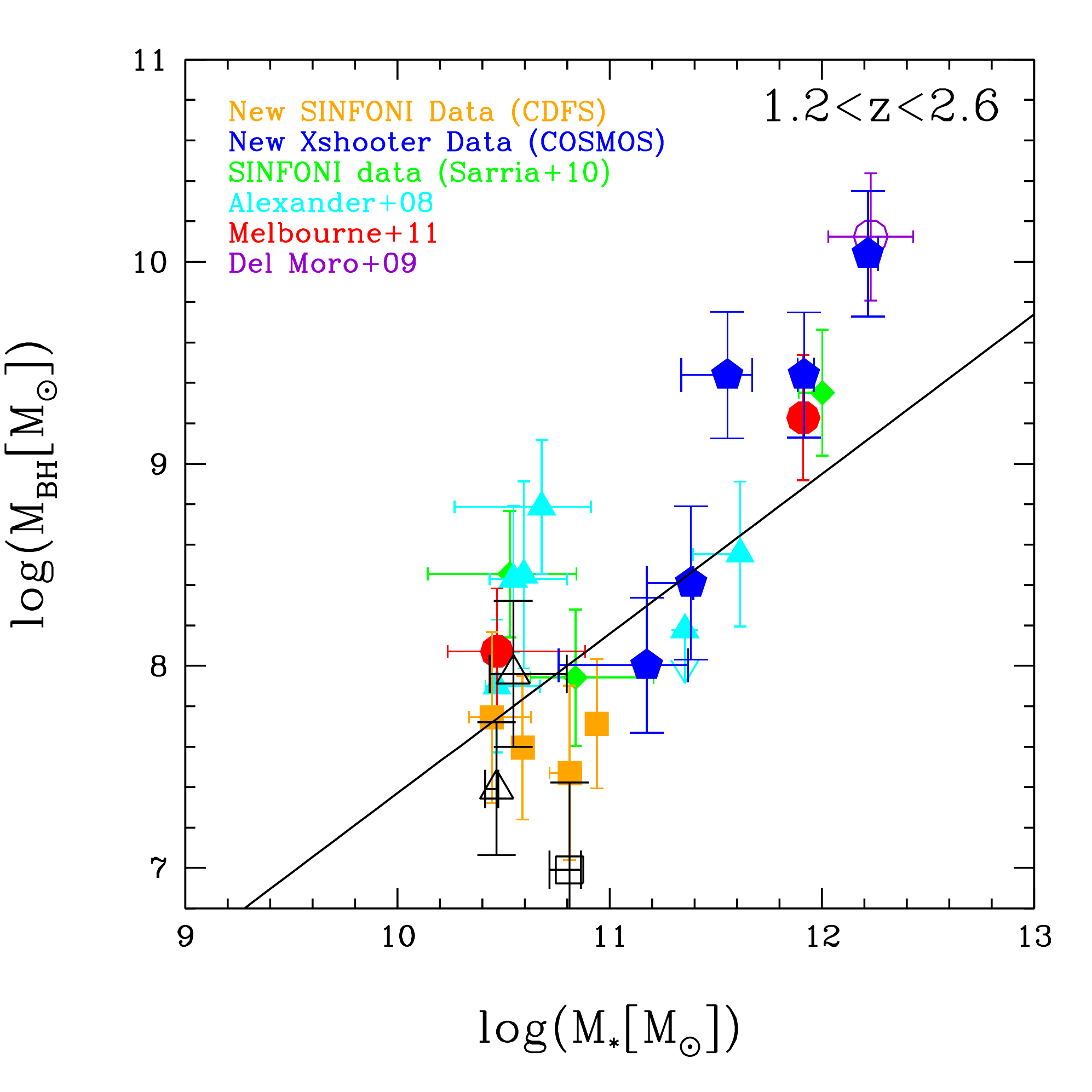}
\caption{Scaling relation  between M$_{\rm BH}$ and M$_{*}$ for obscured AGN at z$\sim$1.2 - 2.6 color-coded as explained in the inset of the plot. The solid line is the local relation from \citet{Sani2011}. 
For the 3 objects (CDFSX633, A1236+6214 and A1237+6203) for which the X-ray luminosity does not agree with the whole SED shape, we show two BH masses: color-coded symbols correspond to the correct SED-based L$_{\rm X}$ while the open symbols (same symbols as the original ones) are the BH masses computed assuming the (lower) L$_{\rm X}$ as measured from X-ray data. } 
\label{fig:scaling}
\end{figure}

In Figure \ref{fig:scaling}, we show the M$_{\rm BH}$ - M$_{*}$ plane for our sample which spans the z$\sim$1.2 - 2.6 range.  
As explained in the previous section, for 3 sources (CDFSX633, A1236+6214 and A1237+6203), the intrinsic X-ray luminosity  disagrees with the rest of the SED. These sources are suspected to be highly obscured (possibly Compton Thick) AGN for which the measured N$_{\rm H}$ is underestimated due to the poor quality of the X-ray data and therefore the intrinsic luminosity much higher than that derived from the X-ray data. Since the intrinsic X-ray luminosity is used to compute the BH mass, the BH masses associated to these objects could be as well underestimated. For these sources, we thus used the SED-derived L$_{\rm X}$ to derive the BH masses. The values for the BH masses considering the L$_{\rm X}$ obtained from the X-ray data are shown in the plot with the same (but open) symbols as the corresponding subsample.

Since the virial relations used in this work to derive the BH masses are normalized to the \citet{Sani2011} local relation (see Sec. 3),  to study the redshift evolution, high redshift sources in this plane have to be compared to the above local relation. 
The \citet{Sani2011} local relation is shown in Fig.  \ref{fig:scaling} with a solid line.  
From this comparison, we find that, X-ray obscured, red QSOs are largely scattered around the local scaling relation. In particular, we note that while less massive objects are equally scattered below and above the local relation, the most massive ones are mainly found above. Considering the Eddington ratios of the analyzed sources, we find both high and low accreting SMBHs equally distributed below and above the local relation.
This is in contrast to the work from \citet{Urrutia2012}, who studied a sample of 13 luminous, red QSOs at z$<$1, and found that low accreting SMBHs ($\lambda<$0.3) lie mainly above the scaling relation while high accreting SMBHs ($\lambda>$0.3) below.

Moreover, we note that the position of some sources (i.e. objects taken from \citet{Sarria2010} and \citet{Alexander2008}) is different compared to what presented in their original papers. This is due to a more accurate estimate of the stellar masses, a different calibration used to compute the BH masses, the use of the L$_{\rm X}$ instead of the L$_{\rm 5100\AA}$ in the BH mass formula for the sources in \citet{Alexander2008} and, for BPM1627\_181, also to a more accurate FWHM measurement. Differences in the BH mass are below 0.4 dex while the host masses can differ up to 0.7 dex.

\begin{figure*}
\includegraphics[width=0.8\textwidth]{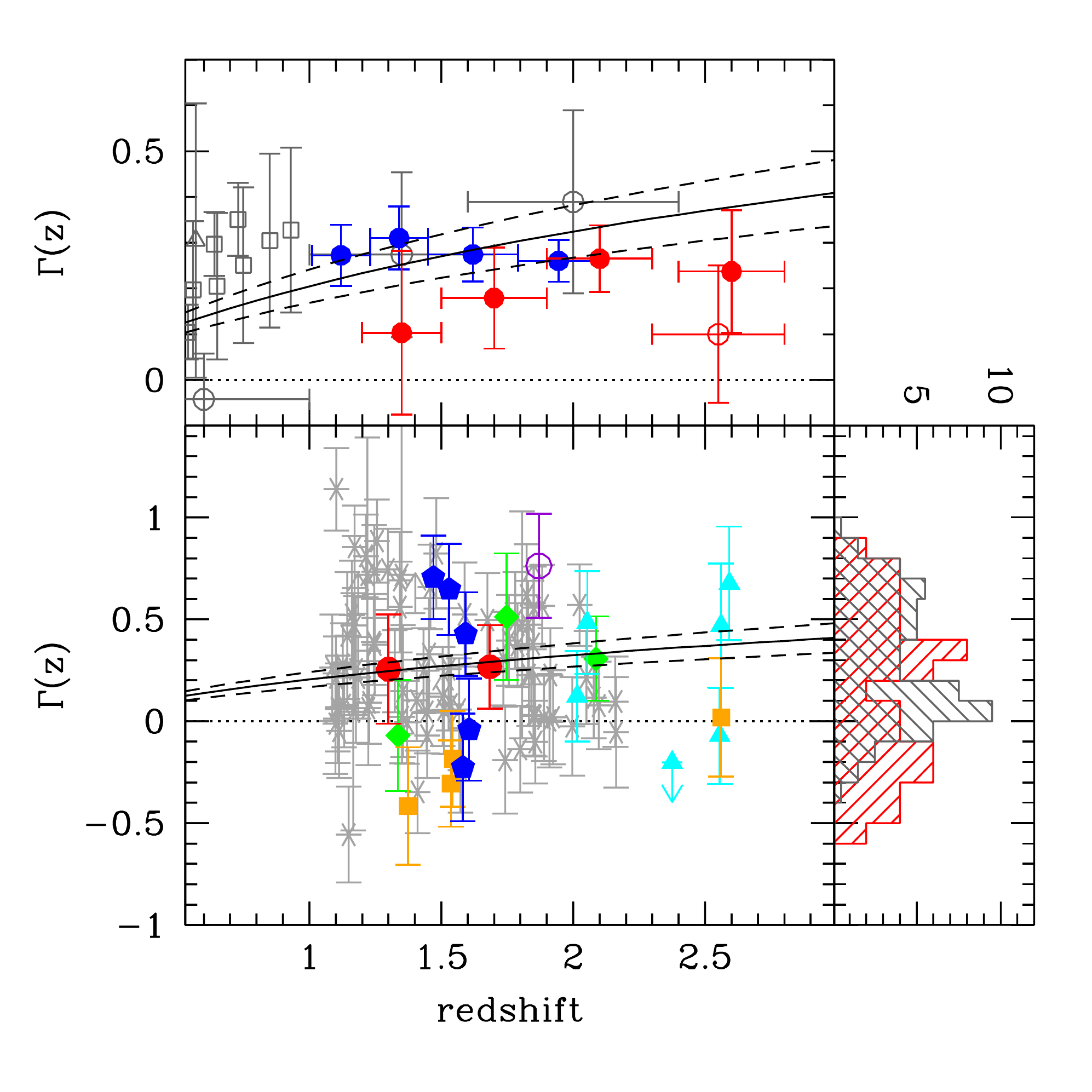}
\caption{\textit{Main panel:} Redshift evolution of the offset measured for our sample from the local M$_{\rm BH}$$-$M$_{*}$ relation. The offset $\Gamma$(z) is calculated as the distance of each point to the \citet{Sani2011} correlation. Different colors and symbols correspond to the different subsamples (as in Fig. \ref{fig:scaling}) while light grey symbols show the comparison sample of type--1, blue QSOs from \citet{Merloni2010}. Solid dark grey line shows the best fit obtained by \citet{Merloni2010} rescaled using an average value to take into account the different formulas used in the BH masses computation. \textit{Upper panel:} red symbols are the average values and errors of our data binned in four redshift bins compared to the blue QSOs from \citet{Merloni2010} (blue symbols). Grey open symbols are objects from literature: triangles are from \citet[][low z]{Salviander2007} and \citet[][high-z]{Shields2003}; squares from \citet{Woo2008} and circles from \citet{Peng2006}. \textit{Right panel:} Histogram of the blue (grey histogram) and red (red shaded histogram) QSOs populations in the common redshift range ($1.2<z<2.0$). For ease of comparison, the red histogram here has been multiplied by 4.} 
\label{fig:scaling_z}
\end{figure*}

\subsection{Evolution of the $\rm M_{BH}-M_*$ scaling relation}

In Fig. \ref{fig:scaling_z}, we plot the offset $\Gamma$(z) between the M$_{\rm BH}$/M$_{*}$ and the local values, calculated as distance in the log space of each point to the \citet{Sani2011} correlation\footnote{Some of the previous works have defined $\Gamma$ as the offset from the scaling relations in terms of ``excess black hole mass''. To compare this measures with our points we have divided these values by $\sqrt{1+A^2}$ where A=0.79 is the slope of the \citet{Sani2011} relation \citep[see also][]{Merloni2010}.} as a function of redshift:

\begin{equation}
\Gamma(z)=log\left(\frac{M_{\rm BH}}{M_{*}}\right)(z)-log\left(\frac{M_{\rm BH}}{M_{*}}\right)(z=0).
\end{equation}

\noindent
In this plot, we also include the comparison sample of optically type--1, blue QSOs at 1.2$<z<$2.1 from \citet{Merloni2010} for which we re-computed the BH masses using eq. \ref{eq:3} of Sec. \ref{sec:BH}, consistently with the dust obscured, red ones.

We see that both  blue and red QSOs show M$_{\rm BH}$/M$_{*}$ ratios increasing going to earlier epochs with red QSOs having slightly lower values at z$<$1.9. This is evident in the upper panel of Fig. \ref{fig:scaling_z} where mean values and errors are shown for both populations. Since the M$_*$ of A1637+4053 is poorly constrained by the SED fit and its M$_{\rm BH}$ is an upper limit due to its L$_{\rm X}$, for the last bin of the red QSOs sample, we show both the mean value taking it into account (open circle) and  excluding it (filled circle). The solid and dashed lines correspond to the redshift evolution derived for blue QSOs at z$<2.1$ and extrapolated at higher redshift \citep{Merloni2010}. High redshift red QSOs do not seem to follow such trend having on average lower values that remain constant at z$>$1.9.  
The two distributions in the common redshift range (1.2$<z<$2.1) are shown in the right panel of Fig. \ref{fig:scaling_z}. According to a Kolmogorov-Smirnov statistics, the probability that the two samples are drawn from the same parent population is 10\%.

\subsection{Possible biases}
\begin{figure}
\includegraphics[width=0.45\textwidth]{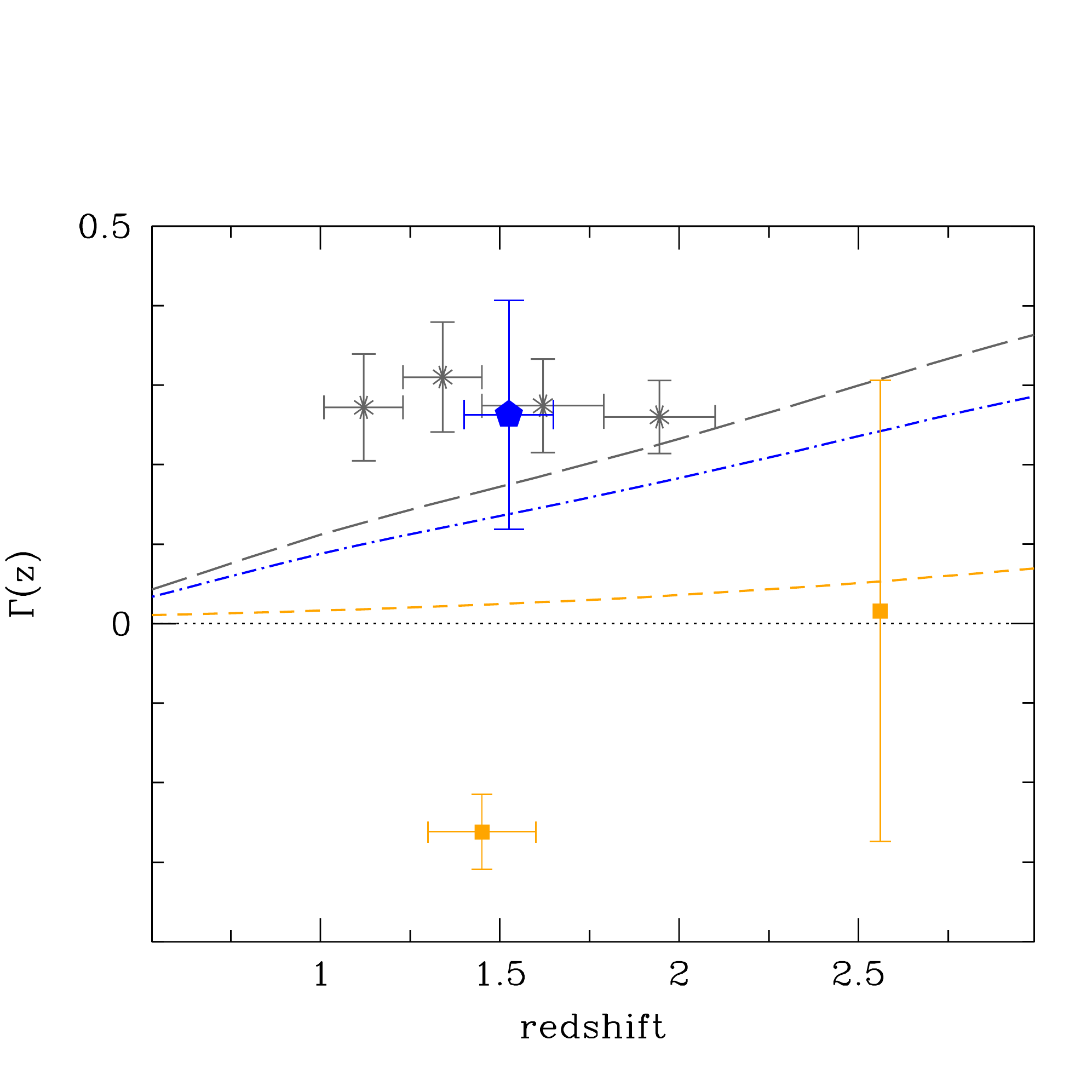}
\caption{Grey circles are optically type-1 blue QSOs as in the upper panel on Fig. \ref{fig:scaling_z} while the average values with their errors for the CDFS-only and the COSMOS-only sources in our sample are shown as orange squares and the blue pentagon, respectively. Note that the orange square at z$\sim$2.5 is made of only one object.
The lines correspond to the expected luminosity bias computed considering the X-ray limiting fluxes of the X-Shooter sample ($\rm 1 \times 10^{-14} erg\,cm^{-2} s^{-1}$; blue dot-dashed line), the Sinfoni sample ($\rm 1 \times 10^{-15} erg\,cm^{-2} s^{-1}$; orange dashed lines) and the XMM-COSMOS sample of optically blue QSOs (long-dashed grey line).} 
\label{fig:scaling_bias}
\end{figure}

As discussed by e.g. \citet{Lauer2007} and \citet{Schulze2011}, AGN flux limited samples are generally biased towards higher values of M$_{\rm BH}$/M$_{*}$. The magnitude of this bias depends on the flux limit of the sample i.e. it increases for brighter limiting fluxes, the underlying distribution functions and the intrinsic scatter in the M$_{\rm BH}$-M$_{*}$ relation. Since both the unobscured AGN sample from \citet{Merloni2010} and the new sample of obscured AGN presented here are flux limited, i.e. X-ray flux limited from XMM-COSMOS and CDFS, both samples are affected by this selection bias. However, since the Chandra Deep Fields observations are very deep, the expected bias is negligible, contrary to the XMM-COSMOS AGN samples for which a non-negligible bias is expected \citep[as already discussed in][]{Schulze2011}. We quantified the magnitude of the bias using the framework presented in \citet{Schulze2011}. We used their Equation~29, assuming the z$=$1 BH mass function and Eddington ratio distribution function derived from the VVDS survey (Schulze et al. in prep) and applying the limiting fluxes corresponding to the considered fields. The result is shown in Fig. \ref{fig:scaling_bias}. In this figure, we show with grey points the averages and their errors of the offsets from the local relation of the unobscured type--1 AGN sample and with orange squares and the blue pentagon the averages for CDFS-only and COSMOS-only obscured sources in the analyzed sample, respectively.

The predicted bias in case of no evolution is shown with a long-dashed grey line for the XMM-COSMOS flux limit. This has to be compared to the grey symbols. For the XShooter and Sinfoni samples we consider as flux limits $\rm f_{X}> 1 \times 10^{-14} erg\,cm^{-2} s^{-1}$ and  $\rm f_{X}> 1 \times 10^{-15} erg\,cm^{-2} s^{-1}$ respectively since above this fluxes the detection rate of broad H$\alpha$ is $>$80\% and in this respect the incompleteness of the sample is negligible (see Sec. \ref{sec:sample_new}). The predicted biases in case of no evolution for such limits are shown as a dot-dashed blue line in the first case and a dashed orange line for the latter case. The dot-dashed blue line has to be compared to the blue point while the dashed orange line to the orange squares.
A slightly different treatment is needed for SMGs and ULIRGs which are primarily selected based on galaxy flux. For both samples, the additional AGN selection criterion, based on the H$\alpha$ flux, leads to a positive bias of the order of $\sim$0.1 dex.
As visible in the plot, the general trend for the sources of our sample as well as for the optical type--1 AGN from Merloni are reproduced by the bias predictions and would lead to the conclusion that both obscured and unobscured AGN lie on the local scaling relation up to z$\sim$2.6. However, the observed samples show a higher mean value of $\Gamma(z)$ (in the case of COSMOS) due to observational biases, as described in \citet{Schulze2011}. 

Interestingly, the CDFS-only mean value at z$\sim$1.5 is much lower than the predicted one given the observational bias. 
In fact, it is important to keep in mind that the estimates of bias effects can suffer from some limitations and uncertainties: 
(i) an accurate prediction of the sample bias at a given redshift requires the knowledge of the underlying distribution functions i.e. the spheroid mass function, the Active BH mass function (BHMF) and the Eddington Ratio Distribution function at that given redshift. Unfortunately, these distributions are still very poorly constrained at high $z$, i.e. the spheroid mass function itself is basically unknown at high $z$. Also the BHMF is best established at the high mass end while the low mass end is less well determined and the systematics not fully understood. Moreover,
(ii) even considering all these uncertainties, a reliable bias prediction requires a sample with a well defined selection function.
This is the case for the optical type--1, blue AGN sample from Merloni which are selected  from the whole XMM-COSMOS sample to be at 1.0$<z<$2.1 and with broad lines in the spectra. The same is not true  for our sources, made of different subsamples of objects for which spectroscopic NIR follow-up was performed.

\section{Evolution of the scaling relation: comparison with the model predictions}
\label{sec:model}

In this section, we compare our observational results with the semi-analytic model from \citet[][and reference therein]{Menci2008} which allows us to follow the accretion onto SMBHs and the related AGN activity together with the evolution of galaxies. 
According to this model, galaxy interactions (i.e. both major and minor merging and fly-by events) trigger both starburst events and SMBH growth by inflow of cold gas in the galaxy disk which is destabilized and thus loses  angular momentum. In addition, a second component of quiescent star formation, corresponding to the gradual conversion of gas into stars on a longer timescale ($1 \rm Gyr$ compared to $\sim$10$^7$yr of the starburst events), is always ongoing and responsible for the galaxy growth. 
The amount of gas available in the system, together with the galaxy interaction rate and their effectiveness in destabilizing the gas, are hence the driving parameters regulating both galaxy and SMBH growths. 
This model correctly describe the tight correlation between the BH mass and the stellar mass in the local Universe \citep{Lamastra2010} and the evolution of the luminous AGN population over a wide range of redshift \citep[see][]{Menci2003,Menci2008}. Such observables are also reproduced by several other semi-analytic models based on the assumption of interaction-triggered AGN implemented with accretion of hot gas and/or of instabilities in self-gravitating disks as additional triggers \citep[see][]{Bower2006,Croton2006,Hopkins2008,Marulli2008,Somerville2008,Guo2011,Fanidakis2012,Hirschmann2012}. 
We compare our observations with the Menci model since it is ideally suited to our purpose. It is indeed the only one that includes a physical description of  AGN feedback providing a distribution of the AGN absorbing column density as a function of the luminosity and redshift tested against observations \citep[see][]{Menci2008}.

By following the merging history of the galaxies at different epochs, it is possible to study the relative growth of BHs and host galaxies and to make predictions on the M$_{\rm BH}$-M$_{\rm *}$ relations at different epochs and as a function of various galaxy properties e.g. mass, gas fraction and star formation rate. 

\citet{Lamastra2010} showed that a typical prediction of such interaction-driven model is that the evolution of the M$_{\rm BH}$-M$_{*}$ relation is stronger for increasing stellar or BH mass. 
Moreover, in agreement with the observations \citep[e.g.][]{Peng2006,Maiolino2007,Merloni2010}, they find that blue QSOs have a higher M$_{\rm BH}$/M$_{*}$ ratio at earlier epochs which decreases going to lower redshift and approaches the local relation.
This trend is due to the fact that, in the early phases, the assembly of BH masses is extremely rapid and much faster than the stellar mass growth since these objects are hosted in the most massive halos formed in dense environments in which both the interaction rate and the fraction of destabilized gas are large. On the contrary, going towards lower redshift, the decline of the interaction rate and of the destabilized gas fraction, suppresses the growth of BHs which is only due to galaxy interactions while quiescent star formation still proceeds continuing to build up stellar mass.

\begin{figure}
\includegraphics[width=0.45\textwidth]{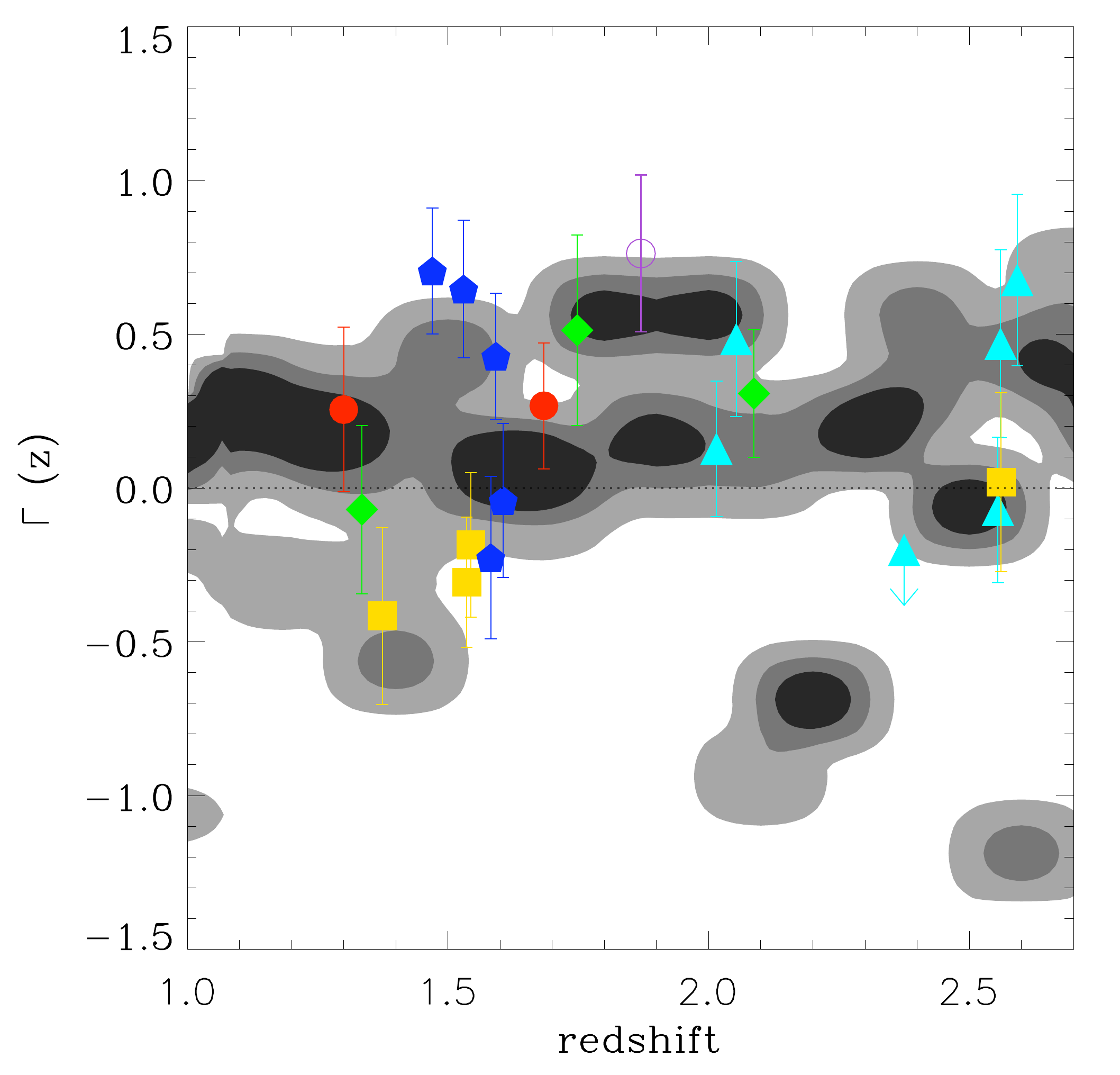}
\caption{Contour plot of the predicted redshift evolution of the offset from the local relation $\Gamma$(z) of N$_{\rm H}>$10$^{22}$ cm$^{-2}$ and $\rm R-K>4$ AGN (\citealt[][and reference therein]{Menci2008}; see also \citealt{Lamastra2010}). The three filled contours correspond to different values of the fraction of objects with a given value of $\Gamma$(z) at the considered redshift: 0.02, 0.1 and 0.2 from the lightest region to the darkest. The colored dots with error bars denote our selected sample with the same coding as in Fig. \ref{fig:scaling}.} 
\label{fig:scaling_z_model}
\end{figure}

In this section, we take a step forward and we follow in the Monte Carlo simulations only the  X-ray obscured, reddened QSO.
The model includes a detailed treatment of AGN feedback which is directly related to the impulsive, luminous AGN phase \citep{Menci2006, Menci2008} and is based on expanding blast waves as a mechanism to propagate outwards the AGN energy injected into the interstellar medium at the center of galaxies. 
In the simulation, AGN are classified as obscured (N$_{\rm H}$ $\ge$ 10$^{22}$ cm$^{-2}$) and unobscured (N$_{\rm H}<$10$^{22}$ cm$^{-2}$) on the basis of the neutral hydrogen column densities corresponding to the unshocked gas in the galaxy disk \citep[equation 6 in][]{Menci2008}.

Since our sample is not a complete sample of sources with a clean selection but rather a collection of sources selected with slightly different criteria, it is not easy to compare them with the model predictions. X-ray obscuration and dust-reddening are common properties of all sources. For this reason, to select in the simulation the sample which is most similar to the data presented in this work, we performed a color cut selecting only X-ray obscured sources with red colors (i.e. R$-$K$>$4 in Vega magnitudes). The simulated galaxy magnitudes are computed from the predicted star formation and chemical enrichment histories using the single stellar population model of \citet{Bruzual2003} and the Salpeter initial mass function. The dust extinction affecting the above magnitudes is computed from the dust optical depth and applying the appropriate attenuation to the luminosity at various wavelengths.
The comparison is shown in Fig. \ref{fig:scaling_z_model} where, together with the observational points (symbols and color-code are as in Fig. \ref{fig:scaling}.), we plot the contour plot of the predicted redshift evolution of the offset from the local M$_{\rm BH}$$-$M$_{*}$ relation ($\Gamma$(z)) for AGN with log(L$_{\rm X})>$43.7, N$_{\rm H}>$10$^{22}$ cm$^{-2}$ and $\rm R-K>4$. These cuts are chosen to best reproduce the selection criteria of the observed sample. The three contour levels correspond to different values of the fraction of objects with a given value of $\Gamma$(z) at the considered redshift: 0.02, 0.1 and 0.2 from the lightest region to the darkest.
   
The model predicts that the $\Gamma(z)$ value for obscured and red QSOs slightly increases towards higher M$_{\rm BH}$/M$_{*}$, with a lower offset than the one  predicted for blue QSOs \citep[see][]{Lamastra2010}. This is in excellent agreement with our observations. We do actually find that most of the sources in our sample lie well inside the contours of the model prediction and that  red X-ray obscured AGN and optical type--1, blue AGN follow a very similar path in the M$_{\rm BH}$-M$_{*}$ plane.

\section{Discussion and Conclusions}
\label{sec:summary}

In this work we have analyzed a sample of X-ray obscured, intermediate type, dust reddened QSOs at 1.5$<z<$2.6. 
Red, obscured QSOs are different from optical type--1 blue QSOs studied so far both in their host and nuclear properties, in that  they are dusty and thus red and their nuclei are obscured. These objects may represent a ``transitional'' phase in the AGN-galaxy co-evolving scenario as expected by the popular theoretical models  by e.g. \citet{Hopkins2008,Menci2004}.  According to these models, in fact, major mergers are the main AGN fueling mechanism and galaxy mergers are able to funnel a large amount of gas onto the nuclear regions in a short time scale.  In this phase the AGN is indeed expected to be obscured and red. Moreover, the resulting high gas density in the central region of the galaxy triggers starburst events: red and obscured QSOs are expected to live in star-forming galaxies as found for most of our sources (90\% are main-sequence or starburst galaxies). Later, as radiative pressure on the dust grains begins to clear the dust and gas away during the blowout phase in which evidence of outflows are expected, the AGN is revealed as an UV-luminous optical type--1, blue QSOs.

We have analyzed the position of these sources in the $\rm M_{BH}-M_*$ plane.
We found a trend with the BH mass, i.e. while less massive red QSOs are about equally scattered below and above the local relation, the most massive ones are mainly located above it. The same trend with BH mass is predicted by the interaction-driven models since high mass BHs form in the most biased regions of the primordial density field where high redshift interactions were favored \citep[see e. g.][]{Lamastra2010}.

Looking at the average M$_{\rm BH}$/M$_*$ ratios as a function of redshift, we find that, similarly to unobscured, optically blue QSOs (e.g. Merloni et al., 2010), obscured red QSOs show an higher ratio compared to the local one, and the increase is higher going to higher redshift (z$\sim$2.6).

Possible observational biases have been analyzed. In particular, as discussed by e.g. \citet{Lauer2007} and \citet{Schulze2011}, AGN flux limited samples are generally biased towards higher values of M$_{\rm BH}$/M$_{*}$. 
Looking at the XMM-COSMOS samples and the CDFS sample, we find that the observed general trend in the $\Gamma(z)-z$ plane is very similar to the one predicted by the bias, thus suggesting that both obscured and unobscured AGN lie on the local scaling relation up to high z but are observed at higher values of $\Gamma(z)$ due to observational biases. \\
However, we must keep in mind that the estimates of bias effects suffer from some limitations i.e.  they require the knowledge of the underlying distribution functions (the spheroid mass function, the active BH mass function and the Eddington ratio distribution function) at the given redshift and unfortunately, these distributions are still very poorly constrained at high $z$.

The computed $\rm M_{BH}/M_*$ average values are consistent with or slightly lower (up to z$<$1.9) compared to  what has been found for blue QSOs \citep{Merloni2010}. 
This result suggests that in the analyzed population of obscured, red QSOs the BH and their host stellar masses are already fully formed and thus their ratio is similar to what observed in optically blue QSOs. 

Recently \citet{Urrutia2012} studying a sample of 13 z$<$1 luminous, dust reddened QSOs, found that these sources are preferentially located below the local scaling relation. This result has been interpreted as the evidence that these sources are in the intermediate phase (blow-out phase, see e.g. \citealt{Hopkins2008}) between the major-merger induced starbursts which appear as ULIRGs and SMGs and the optical type--1, blue QSOs where we expect a rapid BH growth. In our sample, on the contrary, the BH has already grown up and the objects are located on average above the local relation, especially at the high mass end. However, strong outflows are still visible i.e. for the sources for which we have high resolution spectra with good S/N of the [OIII] region from X-Shooter, we indeed find compelling evidence of the presence of outflowing material in the ionized gas component (Brusa et al., 2014).
This suggests that we are observing the final stage of such intermediate phase, when the BH growth is at its end but the AGN feedback has still not finished to clean up the dust and nuclear gas. 
This is not so surprising considering that these sources do show broad lines in their optical lines (redshifted in the NIR): at least part of the obscuring material has already cleared-up and the AGN broad line regions are visible.

Larger samples expanding also to higher redshift are necessary in order to consolidate such results. Large area X-ray surveys with associated multiwavelength follow-up, such as XXL \citep{Pierre2012} and Stripe-82 \citep{LaMassa2013} can be exploited to this end. 
Moreover, IFU spectroscopy and submm observations are fundamental to study in detail such rare objects and to provide a direct observational proof of quasar feedback through the detection of possible galactic scale outflow and on its effect on the host galaxy star formation by mapping the spatial distribution of the SFR \citep[see e.g.][]{Cano-Diaz2012}.

\section*{Acknowledgments}

We thank the referee for the careful reading and the extremely useful comments provided which help to strenght the presented results. This work is based on observations made at the European Southern Observatory, Paranal, Chile (ESO program 88.B-0316(A) and 090.A-0830(A)).
AB work is supported by the INAF-Fellowship Program. MB acknowledge support from the FP7 Career Integration Grant “eEASy” (CIG 321913). 
Support for this publication was provided by the Italian National Institute for Astrophysics (INAF) through PRIN-INAF 2011 “Black hole growth and AGN feedback through the cosmic time” and by the Italian ministry for school, university and research (MIUR) through PRIN- MIUR 2010-2011 “The dark Universe and the cosmic evolution of baryons: from current surveys to Euclid”.
MCD work was supported by the Marie Curie Training Network ELIXIR under the contract PITN-GA-2008-214227 from European Commission and is currently supported by CONACyT (Ciencia Básica) grant 167332-F and by PAPIIT-UNAM grant IA100212.

\bsp
\label{lastpage}

\end{document}